\documentclass{elsart}               
\usepackage{amsmath}
\usepackage{amssymb}
\usepackage{graphicx}
\usepackage{subfigure}
\usepackage{cite}
\usepackage{alltt}
\usepackage{bm}

\begin{document}
\begin{frontmatter}

\title{{\tt GLISSANDO}: {GL}auber {I}nitial-{S}tate {S}imulation {AND} m{O}re...%
\thanksref{grant}} \thanks[grant]{Supported by 
Polish Ministry of Science and Higher Education under grant N202~034~32/0918}
\author[as,ifj]{Wojciech Broniowski},
\ead{Wojciech.Broniowski@ifj.edu.pl}
\author[as]{Maciej Rybczy\'nski},
\ead{Maciej.Rybczynski@pu.kielce.pl}
\author[ifj,rz]{Piotr Bo\.zek},
\ead{Piotr.Bozek@ifj.edu.pl}
\address[as]{Institute of Physics, Jak Kochanowski University, PL-25406~Kielce, Poland} 
\address[ifj]{The H. Niewodnicza\'nski Institute of Nuclear Physics, 
Polish Academy of Sciences, PL-31342 Krak\'ow, Poland}
\address[rz]{Institute of Physics, Rzesz\'ow University, PL-35959 Rzesz\'ow, Poland} 

\begin{abstract}
We present a Monte-Carlo generator for a variety of Glauber-like models (the wounded-nucleon model, binary collisions model, mixed model, 
model with hot spots). These models describe the early stages of relativistic heavy-ion collisions, 
in particular the spatial distribution of the transverse energy deposition which ultimately leads to production of 
particles from the interaction region.
The original geometric distribution of 
sources in the transverse plane 
 can be superimposed with a statistical distribution simulating the 
dispersion in the generated transverse energy in each individual collision.
 The program generates {\em inter alia} the fixed-axes 
(standard) and variable-axes (participant) two-dimensional profiles of the density of sources 
 in the transverse plane and their azimuthal 
Fourier components. These profiles can be used in further analysis of physical 
phenomena, such as the jet quenching, event-by-event hydrodynamics, or analysis of the elliptic flow and its fluctuations. 
Characteristics of the event (multiplicities, eccentricities, Fourier coefficients, {\em etc.}) are stored in a {\tt ROOT} file and can be analyzed off-line. In particular, event-by-event studies can be carried out 
in a simple way. A number of {\tt ROOT} scripts is provided for that purpose. Supplied variants of the code can 
also be used for the proton-nucleus and deuteron-nucleus collisions. 
\end{abstract}

\date{ver. 1.5, 10 February 2008, submitted to Computer Physics Communications}


\end{frontmatter}

\maketitle

\newpage

\noindent{\bf Program summary}

\noindent
{\sl Title of the program:}\-  {\tt GLISSANDO}  \hfill October 2007, ver. 1.5\\
{\sl Catalog identifier:}\- \\
{\sl Program summary URL:}\\ \- http://www.pu.kielce.pl/homepages/mryb/GLISSANDO/index.html \\
{\sl Program obtainable from:}\\ \- http://www.pu.kielce.pl/homepages/mryb/GLISSANDO/index.html \\
{\sl Licensing provisions:}\- none \\
{\sl Computer:} \- any computer with a {\tt C++} compiler and the {\tt ROOT} environment \cite{root},
tested with Intel Core 2 Duo T5200, 1.6~GHz, 1~GB RAM\\
{\sl Operating system under which the program has been tested:} \-
{Linux} Ubuntu~7.04-8.04 ({\tt gcc} 4.1.3-4.2.3), Scientific Linux Linux CERN 3.08 ({\tt gcc} 3.2.3), {Windows XP} with Cygwin ({\tt gcc} 3.4.4 cygwin special), MacBook Pro OSX 10.5.2 ({\tt gcc} 4.0.1),  {\tt ROOT}~ver.~5.08--5.18\\
{\sl Programming language used:} \- {\tt C++} with the {\tt ROOT}
 libraries \\
{\sl Memory required to execute with typical data:} \- below 120~MB\\
{\sl No. of lines in distributed program, including test data:} \- 2000\\
{\sl No. of bytes in distributed program, including test data:} \- 35~kB\\
{\sl Distribution format:} \- tar.gz\\
{\sl Nature of physical problem:} \- Glauber-like models of the initial state in relativistic heavy-ion collisions\\
{\sl Method of solution:} \-  Glauber Monte-Carlo simulation of collision events, analyzed with {\tt ROOT}\\
{\sl Restrictions concerning the complexity of the problem:} \- none\\
{\sl Typical running time:} \- 40~s/10000~events for the wounded-nucleon model and 60~s/10000~events for the mixed 
model with the $\Gamma$ distribution, minimum-bias Au+Au collisions with the fixed-last algorithm for nuclear repulsion, 
dispersion of sources, and hard-sphere wounding profile.  A typical ``physics'' run with 500000 events takes about 1 hour.
The use of the Gaussian wounding profile increases the time about 6
times. (All times for Intel Core~2 Duo T5200 1.6~GHz)\\

\section{Introduction}

The Glauber Monte-Carlo simulations have become a basic tool in the analysis of relativistic heavy-ion
collisions (for a review and the discussion of physics issues see the recent review \cite{Miller:2007ri} and references therein).
The approach provides the initial condition arising just after the collision of two relativistic nuclei.  Within the semiclassical Glauber model, during the first stage of the collision, individual interactions between 
the nucleons deposit transverse energy. 
These elementary processes are classified as {\em wounded nucleons} or {\em binary collisions}.
In this paper the individual deposition of the transverse energy at a definite position in the 
transverse plane is called a {\bf source} and the distribution  of all the sources created in an 
event, weighted by their individual deposited strength, is called the 
{\bf  fireball}. 
A {\bf weight}, called {\bf relative deposited strength} or {\bf RDS}, is assigned to each source when the distribution of
sources in the transverse plane 
is convoluted  with a statistical distribution. This convolution simulates the 
dispersion 
in the generated transverse energy (particle production) from each elementary collision. 
The normalization of the deposited weight can be treated as a additional parameter. This parameter can be separated in the form of an overall factor multiplying the fireball density profile. In the following the deposited weight in each interaction is defined relatively to the one corresponding to an 
elementary nucleon-nucleon 
($NN$) interaction. The RDS strength from an elementary 
interaction encompasses the contribution of two wounded nucleons and one collision and gives on average $1$ (see section\ref{sec:models}).

The Glauber Monte-Carlo methods are used 
to determine the centrality classes in the experiment, since the total nucleus-nucleus cross section is not 
measured at RHIC. Many analysis use the Glauber-model initial state, which can serve as input for cascade simulations 
or hydrodynamics. Moreover, the initial eccentricity of the shape of the fireball
carries over to the elliptic flow coefficient $v_2$ in the momenta of 
produced particles. The behavior of $v_2$ and its fluctuations is a sensitive probe of the dynamics.

Although numerous codes exist for the Glauber Monte-Carlo calculations
and the simplicity of the algorithm allows anybody to produce his own with no difficulty, 
we have decided to make ours publicly available for several reasons.
Firstly, to our knowledge there is no published publicly available package specifically dedicated to 
Glauber Monte-Carlo calculations. Most popular codes are much broader in their scope \cite{Wang:1991hta,Werner:1988jr},
having the initial-state generation only as one of the stages. This makes modifications, extensions, or comparisons 
more difficult. Many details of the model and algorithm implementations are different in these codes, 
such as the choice of the 
nuclear density profiles, the variant of the Glauber model used, the implementation of nucleon-nucleon 
repulsion, {\em etc}. {\tt GLISSANDO} offers flexibility here:
\begin{itemize}
\item There is practically no constraint on the form of the nuclear
density profile. In Sec.~\ref{sec:pos} we discuss the subtleties related to the 
choice of the distribution of the centers of nucleons in the nucleus such that the profiles determined in 
the electron-nucleus scattering are reproduced. 
\item We allow for superimposing a distribution of weights over the distribution of individual sources (wounded nucleons, binary
collisions), since elementary
collisions may result in the deposition of a varying amount of transverse energy. 
\item Hard-sphere or Gaussian wounding profiles may be chosen (see Sect. \ref{sec:coll}).
\item The position of the source may fluctuate relative to the center of the wounded nucleon or 
binary collision from which it originates, which accounts for the finite size of the nucleons. 
\item We offer two different algorithms to simulate the short-range nucleon-nucleon repulsion 
in the generation of the distribution of nuclei in the nucleon.
\end{itemize}
In short, {\tt GLISSANDO} tries 
to encompass all features and details of the used approaches.  Moreover,
\begin{itemize}
\item The {\em variable-geometry} analysis~\cite{Manly:2005zy,Alver:2006wh,Voloshin:2006gz,Broniowski:2007ft}, accounting for the fluctuation of the center of mass and 
the direction of the axes of the moment of inertia of the fireball,
 is built-in in the code. This phenomenon 
is responsible for an increased eccentricity of the initial condition and thus is important in the studies of the 
elliptic flow. 
\item We compute the two-dimensional density profiles 
for both the fixed-axes (standard) 
and variable-axes (participant) geometry, as well as the corresponding Fourier radial profiles. These are stored 
and can be used ``off-line'' in other physics analysis, such as the jet quenching, the initial 
condition for hydrodynamics, {\em etc.} 
\item {\tt GLISSANDO} can also be helpful for studies of event-by-event fluctuations of multiplicities, as measured 
for instance at the CERN
SPS by the NA49 collaboration \cite{ryb}.
\item Basic characteristics of the events are also stored, allowing for event-by-event analysis of various 
quantities.
\item The code can also be used for proton-nucleus and deuteron-nucleus collisions. The Hulthen distribution 
is used to describe the $NN$ separation in the deuteron.
\item The code is written in {\tt C++} and uses the {\tt ROOT} libraries and data structures. That way it is 
tailored for the experimental community. We have taken an effort to document the program such that it can easily be 
modified or extended. 
\item As an option, {\tt GLISSANDO} generates the full event tree containing the positions and RDS of the sources within 
fireballs. This information may be further processed ``off-line'' by other existing codes.
\end{itemize}

We hope {\tt GLISSANDO} will prove useful for heavy-ion physicists.

\section{Basic method}

The simulation of an event consists of three stages: 1)~generation of the positions of nucleons in the 
two nuclei according to the nuclear density distribution, 2)~generation the transverse positions of sources and 
their {\em relative deposited strength} (RDS) -- see Sect.~ \ref{sec:RDS}, 
and 3)~calculation of the event-by-event averaged physical quantities and fireball profiles and 
writing the results to the output file. Stage~3) is performed on-the-fly with stage~2).

\subsection{Generation of positions of nucleons\label{sec:pos}}

The radial nuclear density distribution determined from electron-nucleus scattering has the form
\begin{eqnarray}
n_e(r) = c \frac{4\pi r^2 (1+W_e \frac{r^2}{R_e^2})}{1+\exp(\frac{r-R_e}{a_e})}, \label{wsel}
\end{eqnarray}
where the constant $c$ is such that the normalization 
$\int dr \,n_e(r)=A$ is fulfilled, and the parameters $R_e$, $a_e$, and the Fermi parameter $W_e$ 
can be found in Ref.~\cite{De Jager:1987qc}.
For sufficiently heavy nuclei the radii are well described by the formula
\begin{eqnarray}
R_e=(1.12 A^{1/3} - 0.86 A^{-1/3})~{\rm fm}, \label{rade}  
\end{eqnarray}
while $a_e=0.54$~fm and $W_e=0$, which is the case of the heavy nuclei used at RHIC. From now on we set 
$W_e=0$, although the user may assign any value to this parameter ({\tt WF}) in the input to {\tt GLISSANDO}.

In Glauber Monte-Carlo calculations one generates the positions of {\em centers} of nucleons in the nucleus of mass number $A$.
A relevant point needs to be discussed here \cite{steinberg}.  
The nucleons are not point-like, hence they cause some washing out of the surface.
Therefore the parameters $R$ and $a$ in the distribution of centers of nucleons, $n(r)$, must be such that 
when $n(r)$ is folded with the charge distribution of the single nucleon, $n_N(r)$, the distribution 
$n_e(r)$ is reproduced. We take the nucleon charge profile in the Gaussian form, 
\begin{eqnarray}
n_N(r)=\frac{1}{(2\pi \sigma^2)^{3/2}}\exp \left ( -\frac{r^2}{2 \sigma^2} \right ), 
\end{eqnarray}
where $\sigma=0.79/\sqrt{3}$~fm reproduces the nucleon rms charge radius of 0.79~fm.  
The folding gives 
\begin{eqnarray}
n_e(\vec r)=\int d^3 r' \int d^3 r'' n(\vec r\,') n_N(\vec r\,'') \delta^3(\vec r - \vec r\,' - \vec r\,''), 
\end{eqnarray}
which for the distribution in the radial variable becomes 
\begin{eqnarray}
n_e(r)=2\pi \int_{-1}^1 dz \int_0^\infty  r'^2 dr' n(r') n_N(\sqrt{r'^2+r^2-2 r r' z}). 
\end{eqnarray}
Now, it turns out that the Woods-Saxon form of $n_e(r)$ is well reproduced with a Woods-Saxon form of $n(r)$, albeit 
with different parameters. A fairly good fit is achieved with the formula
\begin{eqnarray}
R&=&(1.12 A^{1/3} - 0.622 A^{-1/3})~{\rm fm}, \label{rad}\\  
a&=&0.46~{\rm fm}, \;\;\;\;({\rm distribution~of~centers}). \nonumber
\end{eqnarray}
Note a shrinkage of the surface thickness parameter $a$ and a slight increase of $R$, with an overall effect of a sharper distribution
of the centers than the electron scattering profile. 

For the convenience of a user wishing to further study these issues, we include a Mathematica script {\tt profile\_folding.nb},
where the folding of the Woods-Saxon profile with the nucleon profile is analyzed. This script is an additional tool and is not 
an integral part of {\tt GLISSANDO}. 

There are two more effects that influence the nuclear density 
distribution in the Monte Carlo approach. The first one is the shift of the generated coordinates to the center-of-mass frame
of the nucleus. This causes some shrinkage of the distribution.
The other effect is the introduction of nucleon-nucleon {\em expulsion}
distance, which is  
a popular way to simulate the short-range repulsion 
in Glauber-like calculations.
The centers of nucleons in each 
nucleus cannot be closer to each other than a certain expulsion distance $d$. The magnitude of $d$ 
should be of the order of the hard-core repulsion range in the nuclear potential. Typically, 
values of a fraction of a fm are used.
{\tt GLISSANDO} offers two ways of generating locations of nucleons satisfying the expulsion constraint. In the first method,
labeled ``fix-last'', the positions are generated subsequently. When the currently generated nucleon is in the forbidden region, {\em i.e.} 
its center is closer than $d$ to a center of any other previously generated nucleons, it is then 
generated anew. The procedure is repeated, until 
a ``good'' location is found. In the second method, labeled ``return-to-beginning'', when a nucleon is generated in the forbidden region  
the whole sequence of locations of the previously found nucleons 
is discarded and the procedure is started from the beginning, until a ``good'' sequence of nucleon locations is formed. The two
methods yield virtually the same results for all investigated observables, and since the ``fix-last'' method is a few 
times faster (with $d=0.4$~fm), it is preferred. Nevertheless, the user has a choice between the two generation methods.  The value $d=0.4$~fm is a typically range of the repulsive core in the $NN$ interaction and is used as default in our input  files.

The repulsion implemented 
via the condition $d>0$ increases somewhat (at the level of 1\%) the size $R$ of the nucleus, 
which is noticeable in both above-described methods.
This effect, and the previously mentioned shrinkage due to the shift to the center of mass, must be compensated by changing appropriately 
the parameters of the ``bare'' distribution from which the positions of centers of nucleons are generated.
This is important, as otherwise a distribution different from the desired one will be effectively obtained.
We find that the following parameterizations do a fairly good job (here we give them for the "fixed-last" method
only):
\begin{eqnarray}
R&=&(1.114 A^{1/3} - 0.246 A^{-1/3})~{\rm fm}, \label{rad0}\\  
a&=&0.45~{\rm fm}  \;\;\;\;\;\; (d=0~{\rm fm} - {\rm no~expulsion}),\nonumber \\\nonumber \\
R&=&(1.113 A^{1/3} - 0.277 A^{-1/3})~{\rm fm}, \label{rad04}\\  
a&=&0.45~{\rm fm} \;\;\;\;\;\; (d=0.4~{\rm fm}), \nonumber \\\nonumber \\
R&=&(1.103 A^{1/3} - 0.550 A^{-1/3})~{\rm fm}, \label{rad08}\\  
a&=&0.455~{\rm fm} \;\;\;\;\;\; (d=0.8~{\rm fm}). \nonumber 
\end{eqnarray}
These formulas are implemented in {\tt GLISSANDO}. 
In order to use them, one needs to set {\tt RWSA=-1} and {\tt RWSB=-1} in the input file. The parameterizations are provided 
only for the cases $d=0$, $0.4$, or $0.8$~fm and for {\tt RET=0} (fixed-last method). 
Otherwise the user must explicitly give the values for the distribution parameters
in the input file.

We summarize: parameters (\ref{rad0}-\ref{rad08}) produce approximately the Woods-Saxon 
distributions of centers of nucleons with parameters (\ref{rad}), which upon folding with the finite-size charge distribution of the
nucleon reproduce to a very good approximation formula (\ref{wsel}) with parameters (\ref{rade}). 

The user can override the default parameters in the input file, as explained below, as well as examine the resulting distribution 
of centers of nucleons. The functional form of the distribution may be changed 
straightforwardly by modifying the formulas in {\tt functions.cpp}. 

For the special case of the deuteron we use the Hulthen 
distribution \cite{hulten} for the distance between 
the centers of the two nucleons,
\begin{eqnarray}
n_d(r)=\frac{2a_d b_d(a_d+b_d)}{(a_d-b_d)^2}\left [ \exp(-2a_dr)+\exp(-2b_dr)-2\exp(-(a_d+b_d)r) \right ], \nonumber \\ \label{hulthen}
\end{eqnarray}
with $a_d=0.228~{\rm fm}^{-1}$ and $b_d=1.118~{\rm fm}^{-1}$.

\subsection{Collision \label{sec:coll}}

After generating the two nuclei, their centers-of-masses must be separated by the impact parameter ${\bm b}$. Technically,
the $(x,y)$ coordinates of the nucleons in nucleus $A$ are shifted by a fixed transverse vector such that
their center-of-mass position is $(b B/(A+B),0)$, while the nucleons in 
nucleus $B$ are shifted such that their center-of-mass is at $(-b A/(A+B),0)$. 
In this frame we call the point $(0,0$ the {\em geometric center of mass}.
Next, the wounded nucleons and the binary collisions are counted. 

\subsubsection{Hard-sphere wounding \label{sec:hard}}
In the hard-sphere wounding prescription, a nucleon from one nucleus 
is wounded if its center passes closer (in the transverse plane) to the center of any of the nucleons from the second nucleus than the wounding distance
\begin{eqnarray}
r_0=\sqrt{\sigma_{w}/\pi}, \label{r0} 
\end{eqnarray}
where $\sigma_{w}$ is equal to the $NN$ inelastic cross section at a given collision energy.
The standard implementation of the wounded nucleon model at 
the RHIC $\sqrt{s_{NN}}=200$~GeV
energy assumes that the inelastic cross-section in 
the nucleon-nucleon collision is \cite{pdg}
\begin{eqnarray}
\sigma_{w}=42~{\rm mb}. \label{swound}
\end{eqnarray}
The cooresponding value for the LHC energies of $\sqrt{s_{NN}}=5500$~GeV is $\sigma_{w} \simeq 63~{\rm mb}$, while for the SPS 
energies of $\sqrt{s_{NN}}=19$~GeV one uses $\sigma_{w} \simeq 32~{\rm mb}$.

Similarly, a binary collision occurs if centers of two 
nucleons, the  first one  from the nucleus
$A$ and the second from $B$, pass closer to each other than the distance
$r_0$. This means that in this model binary collisions are counted 
as if occuring with the same inelastic cross-section as for 
wounded nucleons.

\subsubsection{Gaussian wounding \label{sec:gauss}}
Instead of the hard-sphere method, one may use a smooth function to determine the probability distribution of wounding or a binary
collision. An option 
in {\tt GLISSANDO} allows the user to use the Gaussian wounding function, 
\begin{eqnarray}
p(r)= G \exp \left ( -\frac{G r^2}{r_0^2} \right ),
\end{eqnarray}
where $r$ is the transverse distance between the centers of nucleons and
the normalization is $\int_0^\infty 2\pi r p(r) dr = \pi r_0^2=\sigma_w$. 
The parameter $G$ controls the central value of the profile. 
The default value is $G=0.92$, which is taken from the experimental $pp$ analysis of \cite{Amaldi:1979kd,Bialas:2006qf}. 
In order to switch on this feature the 
preprocessor option {\tt -D\_gauss\_=1} must be used in the makefiles. 

\subsubsection{Displacement of sources}
Since the nucleons have a finite size, the location of the source may be somewhat displaced 
relative to  the center of the wounded nucleon or the mean of the centers of the two nucleons 
for the case of the binary collisions.
{\tt GLISSANDO} may include this effect by randomly displacing the $x$ and $y$ coordinates with a Gaussian distribution 
of width {\tt DWS} for the wounded nucleons and {\tt DBIN} for the binary collisions.  
Non-zero values of these parameters increase somewhat the size of the fireball, as well as reduce
the eccentricity. This has also been noticed in later studies of Ref.~\cite{Alver:2007cs}. 
Although the displacement effect is very physical, it is non-standard, hence is not included in the default parameters. 
If the user wishes to put it in, the 
typical values for the parameters {\tt DW} and {\tt DBIN} should be around 0.7~fm, a typical nucleon size.

\section{Models\label{sec:models}}

In this section we describe the models implemented in {\tt GLISSANDO}.

\subsection{The relative deposited strength \label{sec:RDS}}

The collisions between nucleons (wounding or binary collisions) result in deposition of a certain amount of 
the transverse energy (or entropy) at the location of the source in the
transverse plane, which is then  carried away 
by the produced particles. This quantity
may be different for the wounded nucleons and the binary collisions, as well as can 
fluctuate from source to source. For the studies of fluctuations the absolute normalization 
of the strength of the sources is irrelevant, hence we use the measure proportional to the 
deposited transverse energy, the {\bf relative deposited strength} (RDS).
For the wounded nucleon model RDS is just half of the number of wounded nucleons, which is a convention. For more involved models 
it is composed of the number of wounded nucleons and binary collisions, possibly with an overlayed statistical distribution 
of strength (see the following). 

\subsection{Wounded nucleon model}

In the wounded-nucleon model
the RDS of $1/2$ is attributed to each point in the 
transverse plane at the center of a wounded nucleons.
This complies to the original convention of Ref.~\cite{Bialas:1976ed}, where the average multiplicity of produced 
particles is $N_{\rm part}=N_{pp} N_w/2$, with $N_{pp}$ denoting the average multiplicity in proton-proton
collisions.   

\subsection{Binary collisions}

For binary collisions the RDS of $1$ is attributed to each collision point, 
which is taken as the mean of the transverse coordinates of the 
two colliding nucleons.

\subsection{Mixed model}

A successful description of multiplicities at RHIC has been achieved with a {\em mixed} model \cite{Kharzeev:2000ph}, 
amending the wounded nucleon model  \cite{Bialas:1976ed}
with some binary collisions \cite{Back:2001xy,Back:2004dy}. 
In this case a wounded nucleon obtains the RDS of \mbox{$(1-\alpha)/2$}, and a binary collision 
the RDS of $\alpha$. The total RDS averaged over events 
is then $(1-\alpha)N_{\rm w}/2+\alpha N_{\rm bin}$. The fits to particle 
multiplicities of Ref.~\cite{Back:2004dy} give $\alpha = 0.145$ for
collisions at $\sqrt{s_{NN}}=200$~GeV, and $\alpha = 0.12$ for $\sqrt{s_{NN}}=19.6$~GeV.

Thus the $\alpha$ ({\tt ALPHA}) parameter in the code controls the proportion of wounded nucleons to binary collisions, with 
$\alpha=0$ corresponding to the pure wounded-nucleon model, $\alpha=1$ to binary collisions only, and 
the intermediate values to the mixed model. 

\subsection{The hot-spot model}

We also implement a model with {\em hot spots} in the spirit of Ref.~\cite{Gyulassy:1996br}, assuming that the 
cross section for a semi-hard binary collision producing a hot spot 
is small, $\sigma_{\rm bin} \simeq 2$~mb, which is a parameter of the model. Importantly, when such a rare collision occurs it produces 
on the average a large amount of transverse energy, with RDS equal to \mbox{$\alpha\sigma_{\rm w}/\sigma_{\rm bin}$}.
Technically, the hot spots are implemented as follows: the binary collision is generated according to the criterion
of Sect.~\ref{sec:coll}, but is accepted with the probability
 \mbox{$\sigma_{\rm bin}/\sigma_{\rm w}$}, {\em i.e.} 
the RDS of $\alpha \sigma_{\rm w}/\sigma_{\rm bin}$ is assigned to the hot spot at its position in the transverse plane. 
Otherwise the RDS is zero.  
All the wounded nucleons have the RDS assigned to 
$(1-\alpha)/2$.
In this paper we label any model with $\sigma_{\rm bin}<\sigma_{\rm w}$ the hot-spot model, although 
it should actually mean $\sigma_{\rm bin} \ll \sigma_{\rm w}$.
The weight an an event is on average equal to 
$(1-\alpha)N_{\rm w}/2+\alpha N_{\rm bin}$, the same as in the mixed model, but
it can fluctuate considerably from event to event depending 
on how many hot spots are created.

\subsection{Superposition}

Each source from the previously described models (wounded-nucleon, binary, mixed, or hot-spot) may 
deposit the transverse energy with a certain probability distribution. To incorporate this effect, we
superimpose a statistical distribution of weights over the distribution of sources. {\tt GLISSANDO}
has three options here, controlled by the parameter {\tt MODEL}: 0~-~no superposition 
(all superimposed weights equal to one), 1~-~superposition of the Poisson 
distribution, and 2~-~superposition of the gamma distribution. 

The Poisson distribution 
generates the discrete weights according to the formula
\begin{eqnarray}
g(w,\kappa )=\frac{\kappa ^{w \kappa} \exp(-\kappa )}{(w \kappa)!}, \;\; w=0,\frac{1}{\kappa}, \frac{2}{\kappa}, 
\frac{3}{\kappa}, \dots \label{poiss}
\end{eqnarray} 
This distribution has $\langle w \rangle=1$ and $\sigma(w)=1/\sqrt{\kappa}$. 

\begin{figure}[t]
\begin{center}
\subfigure{\includegraphics[width=.7\textwidth]{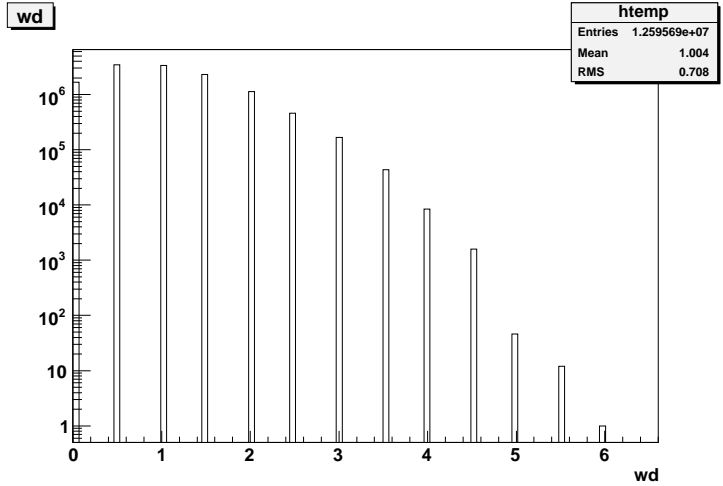}}
\end{center}
\caption{Histograms for the superposition distribution generated by {\tt GLISSANDO} from the Poisson 
distribution, $\kappa=2$, 20000~events. The figure is generated by running {\tt ./glissando\_profile.exe input\_profile\_0.dat} 
and then using the {\tt wd} branch in the {\tt density} tree of the output file 
{\tt glissando.root}. Horizontal axis: weight, vertical axis - probability distribution. Normalization arbitrary. Plot generated with {\tt ROOT}. \label{fig:supdis}}
\end{figure}

The $\Gamma$ distribution generates continuous weights according to the density 
\begin{eqnarray}
g(w,\kappa)=\frac{w^{\kappa-1}\kappa^\kappa \exp(-\kappa w)}{\Gamma(\kappa)}, \;\; w \in [0,\infty). \label{gamma}
\end{eqnarray} 
This distribution also gives $\langle w \rangle=1$ and $\sigma(w)=1/\sqrt{\kappa}$.

The superposition distributions can be supplied independently for the wounded nucleons and binary collisions. 
The parameters $\kappa$ for both distributions are denoted as {\tt Uw} and {\tt Ubin}, respectively. 

The preprocessor variable {\tt \_weight\_} controls the output of the superposition density to the output {\tt ROOT}
tree. The binary {\tt glissando\_profile.exe} writes out the superposition weights for the
wounded nucleons.

The fact that the chosen superposition distributions have $\langle w \rangle=1$ is at no loss of generality, 
since, as already stated, the individual RDS used for carrying the statistical averages 
can be normalized arbitrarily. The average RDS per event is still 
$(1-\alpha)N_{\rm w}/2+\alpha N_{\rm bin}$ and does  not
 change due to the superposition procedure.

For the general case we assign the RDS of 
$(1-\alpha) w/2$ for the wounded nucleons and  
$\alpha w \sigma_{\rm w}/\sigma_{\rm bin}$ 
for those binary collisions which are accepted with the probability
 $\sigma_{\rm bin}/\sigma_w$. The variable $w$ is generated with one of the distributions described above. 
The wounded-nucleon model corresponds to
 $\alpha=0$, the mixed model to $\alpha>0$ and
$\sigma_{\rm bin}=\sigma_{\rm w}$, and the hot-spot model to $\alpha>0$ and
$\sigma_{\rm bin}<\sigma_{\rm w}$.

\section{Fixed- and variable-axes quantities}

We use the standard convention for the axes of the reference frame: the $z$-axis is along the beam, the $x$-axis 
lies in the reaction plane,
and the $y$-axis is perpendicular to the reaction plane. The azimuthal angle $\phi \in [-\pi,\pi]$
is measured relative to the $y$-axis, hence 
\begin{eqnarray}
y=\rho \cos \phi,  \;\; x=\rho \sin \phi, 
\end{eqnarray}
where $\rho=\sqrt{x^2+y^2}$ is the transverse radius. 

We refer to the analysis in the reference frame  fixed by the reaction plane as {\bf fixed-axes} (also called {\em standard} in 
the literature), and to the analysis where the particles in each event are translated to the center-of-mass frame and aligned with the 
major principal axis of the second harmonic moment of the fireball as {\bf variable-axes} (also called {\em participant}).  

If the reaction plane is determined in each event (which of course can never be achieved 
precisely in the experiment \cite{Poskanzer:1998yz,Ollitrault:1992bk}), one can 
choose the reference frame fixed by the reaction plane. 
The boost-invariant (in the central rapidity region) 
two-dimensional  profile of the fireball density,
 $f(\rho,\phi)$, 
is obtained by averaging the RDS distribution over the events belonging to a particular centrality class.  
Thus the normalization is $\int d\phi \rho d\rho f(\rho,\phi)=$~average~total~RDS. For instance, in the wounded nucleon model 
$\int d\phi \rho d\rho f(\rho,\phi)=N_w/2$.
The symmetry $f(\rho,\phi)=f(\rho,\pi-\phi)$, occurring also for unequal nuclei, excludes odd cosine and even sine components in the Fourier decomposition, 
hence in the general case of unequal nuclei we have 
\begin{eqnarray}
f(\rho,\phi)&=&f_0(\rho) + 2 f_2(\rho) \cos(2\phi)+ 2 f_4(\rho) \cos(4\phi)+ \dots \nonumber \\
&+&2 g_1(\rho) \sin(\phi) +2 g_3(\rho) \sin(3\phi)  + \dots  \;\;\;\;\;\;\;\;\; (A\neq B).\label{fg}
\end{eqnarray}
For equal colliding nuclei the symmetry $f(\rho,\phi)=f(\rho,-\phi)$ eliminates the sine functions,
\begin{eqnarray}
f(\rho,\phi)=f_0(\rho) + 2 f_2(\rho) \cos(2\phi)+ 2 f_4(\rho) \cos(4\phi)+ \dots, \;\; (A=B). \label{f}
\end{eqnarray}
If the switch {\tt SHIFT} is set to 0, 
the fixed-axes distance $\rho=\sqrt{x^2+y^2}$ is measured from the center of the geometric overlap of the two nuclei, which is the 
traditional way.
For equal nuclei it is just the mid-point between their centers of mass. 
If {\tt SHIFT=1}, then $\rho$ is computed in each event 
relative to the center of mass 
of the fireball,
{\em i.e.}, the center of mass of the RDS deposited in that event. 
The harmonic moments obtained from (\ref{fg},\ref{f}), which we call {\em fixed-axes}, are also called ``standard'' 
in the literature.

In order to have some convenient quantitative measures of the profiles of Eq.~(\ref{f})
one introduces their radial moments
\begin{eqnarray}
\epsilon_{k,l}=
\frac{\int 2\pi \rho f_l(\rho) \rho^k d\rho }{\int 2\pi \rho f_0(\rho) \rho^k
d\rho }. \label{epsilon}
\end{eqnarray}
The choice of the weighting power $k$ is arbitrary, with the typical choice $k=2$. 
Higher values of $k$ would make the measure more sensitive to the outer region of the system, see
Ref.~\cite{Broniowski:2007ft} for a discussion of this point. We note that in the popular notation 
\begin{eqnarray}
\epsilon_{\rm std}=\epsilon_{2,2}\equiv \epsilon.
\end{eqnarray} 

Experimentally, the reaction plane cannot be determined accurately. Numerous methods have been developed to analyze
the azimuthal asymmetry in heavy-ion collisions. The role of the purely statistical fluctuations caused 
by the finite number of particles has been described in 
 \cite{Aguiar:2000hw,Aguiar:2001ac,Miller:2003kd,Bhalerao:2005mm,Andrade:2006yh,Voloshin:2006gz,Broniowski:2007ft,Voloshin:2007pc,%
Manly:2005zy,Alver:2006wh,Alver:2007cs}.
As shown in these papers, the effects of the {\em variable geometry} in the initial stage of the 
collision lead to important effects, in particular to an
increase of the shape eccentricity resulting in increased values of the elliptic flow coefficient $v_2$. 

In the variable-axes calculations the coordinates of the elementary 
sources in the given event are first 
shifted to the center-of-mass of the fireball.
One then computes the principal axes of the ellipse of inertia of the fireball generated in the event, which is twisted relative to the reaction plane 
coordinate system. 
The angle between the major half-axis of the ellipse and the $y$ 
axis is given by the relation
\begin{eqnarray}
\tan (2\phi^\ast) = 2 \frac{\langle xy \rangle - \langle x \rangle \langle y \rangle}{{\rm var}(y)-{\rm var}(x)}, \label{tanast}
\end{eqnarray}  
where var denotes the variance.
The brackets denote the average over the sources in the given event. 
The angle $\phi^\ast$ fluctuates from event to event.
The superscript $\ast$ indicates quantities averaged in such a way, that first in each event one computes from Eq.~(\ref{tanast})
the rotation angle $\phi^\ast$, then the rotation is performed to 
the current principal-axis system, and finally averaging over events is performed. The procedure results in the 
``variable-axes'' density 
\begin{eqnarray}
f^\ast(\rho,\phi)&=&f^\ast_0(\rho) + 2 f^\ast_2(\rho) \cos(2\phi-2\phi^\ast) 
+ 2 f^\ast_4(\rho) \cos(4\phi-4\phi^\ast)+ \dots \label{fgr} \nonumber \\
&+&2 g_1^\ast(\rho) \sin(\phi-\phi^\ast) +2 g_3^\ast(\rho) \sin(3\phi-3\phi^\ast)  + \dots  \;\;\; (A\neq B),
\end{eqnarray}
or for the case of equal nuclei 
\begin{eqnarray}
f^\ast(\rho,\phi)&=&f^\ast_0(\rho) + 2 f^\ast_2(\rho) \cos(2\phi-2\phi^\ast)  
+ 2 f^\ast_4(\rho) \cos(4\phi-4\phi^\ast)+ \dots \nonumber \\ && \hspace{8.7cm} (A = B). \label{fr} 
\end{eqnarray}
We refer to the above density profiles as to the {\em variable-axes} profiles.

In analogy to Eq.~(\ref{epsilon}) we introduce the variable-axes moments
\begin{eqnarray}
\epsilon^\ast_{k,l}=\frac{\int 2\pi \rho f^\ast_l(\rho) \rho^k \, d\rho}{\int 2\pi \rho f^\ast_0(\rho) \rho^k \, d\rho}, \label{epsilona}
\end{eqnarray}
In the commonly-used notation for the variable-axes or {\em participant} deformation parameter one has
\begin{eqnarray}
\epsilon_{\rm part}=\epsilon^\ast_{2,2} \equiv \epsilon^\ast. \label{epart} 
\end{eqnarray}
The profiles and moments for higher harmonics are suppressed, similarly to the fixed-axes case. 
This is clear, as the higher harmonics are evaluated relative to the axes determined by maximizing 
the quadrupole moment. As a result, only a few moments are needed to effectively parameterize the profile.

The shift to the center-of-mass of the fireball results in a slightly more compact distribution $f^{\ast}_0(\rho)$ than
$f_0(\rho)$ in the geometric center-of-mass frame, see Fig.~\ref{fig:prof}
 below.

In Ref.~\cite{Broniowski:2007ft} we show that many of the qualitative and quantitative features of the 
Fourier distributions displayed above result from purely statistical considerations.  

{\tt GLISSANDO} generates and stores the fixed- and variable-axes moments, as well as the two-dimensional 
density profiles of the fireball and their harmonic decomposition. 
They can be later used in other 
studies involving the shape of the medium, such as jet quenching \cite{Broniowski:2007ft}, setting the initial condition for 
cascade or hydrodynamic studies, {\em etc.} 
The included program {\tt interpolation} shows a sample use of the profiles in a {\tt C++} code. 

\section{Typical sequence of running}

\subsection{Installation}

It is necessary to have the {\tt ROOT} package installed.
After obtaining the {\tt GLISSANDO} distribution the user should unpack

\begin{verbatim}
tar -xzvf glissando_1_5.tar.gz
\end{verbatim}

and run the installation script

\begin{verbatim}
sh install
\end{verbatim}

\noindent
As a result, executable binaries are created. Five of them 
perform simulations:  {\tt glissando.exe} for the nucleus-nucleus (A+B) collisions,  {\tt glissando\_prot.exe} for the 
proton-nucleus (p+A) collisions, {\tt glissando\_deut.exe} for 
the deuteron-nucleus (d+A) collisions, and also {\tt glissando\_profile.exe}, 
which in addition generates the nuclear profile of nucleus $A$ while running the (A+B) collisions. All these 
codes implement the hard-sphere wounding profile, see Sect.~\ref{sec:hard}. The binary  {\tt glissando\_gauss.exe} 
performs the (A+B) collisions with the Gaussian wounding profile, see Sect.~\ref{sec:gauss}. The wounding profile is controlled with the 
{\tt \_gauss\_} flag in the makefiles ({\tt -D\_gauss\_=0} for hard sphere, {\tt -D\_gauss\_=1} for 
Gaussian), hence the user may easily implement his choice also in the codes for the (p+A) and (d+A) collisions.
The code {\tt interpolation.exe}
provides an example of creating interpolating functions of the two-dimensional density profiles and their Fourier-decompositions, 
based on the stored histograms. The code {\tt retrieve.exe} illustrates a use of the full event tree, optionally 
generated by {\tt GLISSANDO}. 
Several useful {\tt ROOT} scripts are described in Appendix~\ref{scripts}.

\subsection{Running the simulations \label{sec:run}}

\begin{table}[tb]
\caption{Typical output to the console from {\tt GLISSANDO}.\label{tab:out}}
\begin{center}
{\small
\begin{verbatim}
********************************************************
GLISSANDO ver. 1.6 ( http://arxiv.org.abs/0710.5731 )
               tested with ROOT ver. 5.08--5.18
********************************************************
Start: Tue Feb  5 20:47:46 2008
--------------------------------------
  nucleus - nucleus collisions
--------------------------------------
parameters reset in input file input.dat :
EVENTS  100000
ALPHA   0
BMAX    20

generates output file glissando.root
random seed: 1202240866, number of events: 100000
197+197, RA=6.43, aA=0.45, RB=6.43, aB=0.45, d=0.4fm
wounded nucleon model: sig_w=42mb
   (binary collisions not counted)
hard sphere NN collision profile
window: b_min=0fm, b_max=20fm
fix-last algorithm

event: 100000     (100%)

Quantities for the specified b, N_w, and RDS window:
A+B cross section = 6436.6mb, equiv. hard-sphere radius = 7.15687fm
efficiency (accepted/all) = 51.2208%
N_w = 103.179+/-105.16
RDS = 51.5897+/-52.58
eps_fix_2  (std.) = 0.315127+/-0.316701
eps_var_2 (part.) = 0.475002+/-0.275256
eps_fix_4 = 0.0872776+/-0.281066
eps_var_4 = 0.240773+/-0.31374
x_cm = -0.000206385+/-0.482281,  y_cm = 0.00400375+/-0.732762

Finish: Tue Feb  5 20:54:11 2008
(0h:6m:25s)
**************************************
\end{verbatim}
}
\end{center}
\end{table}
For simulations of the A+B collisions the running command has the syntax

\begin{verbatim}
./glissando.exe [input_file] [output_file]
\end{verbatim}

and analogously for the p+A ({\tt glissando\_prot.exe}) and d+A ({\tt glissando\_deut.exe}) cases.
When the input and output file-name arguments are absent, their default values are 
\begin{verbatim}
input.dat - default input
glissando.root - default output
\end{verbatim}
Typical input files are also provided with the distribution. The input 
parameters and their defaults are described in Appendix~\ref{input}.
Thus we may simply type
\begin{verbatim}
./glissando.exe
\end{verbatim}

\noindent 
The output to the console shown in Table~\ref{tab:out} contains basic information on the run. 
The subsequent lines give the info on the input: the version of the code, initial time, type of reaction, name of the input file used and the
values of parameters reset from the default, the seed for the {\tt ROOT} 
random-number generator, the number of events, the mass numbers
of nuclei, the ``bare'' (see Sect.~\ref{sec:pos}) 
Woods-Saxon parameters and the expulsion distance, the type and parameters of the model (wounded, binary, 
mixed, hot-spot), the window in the impact parameter (and the number of wounded nucleons and the value of total weight), 
the type of the algorithm for the 
short-range repulsion, the dispersion parameters for the location of sources, and the counter for events. The final output consists of the 
total nucleus-nucleus cross section in the given window, $\sigma_{AB}$, the equivalent hard-sphere radius defined as $\sqrt{\sigma_{AB}/\pi}/2$, and the efficiency parameter, denoting the ratio of events where the nuclei collided to all 
the Monte-Carlo generated events. Next come averages of basic quantities with their standard deviations: the number of the wounded nucleons, 
binary collisions, hot spots, RDS, eccentricity $\epsilon$ in the fixed- and variable-axes frame, $\epsilon_4$ 
in the fixed- and variable-axes frame, the center-of-mass of the fireball. Finally, the execution time is printed.

\begin{figure}[tb]
\begin{center}
\subfigure{\includegraphics[width=.7\textwidth]{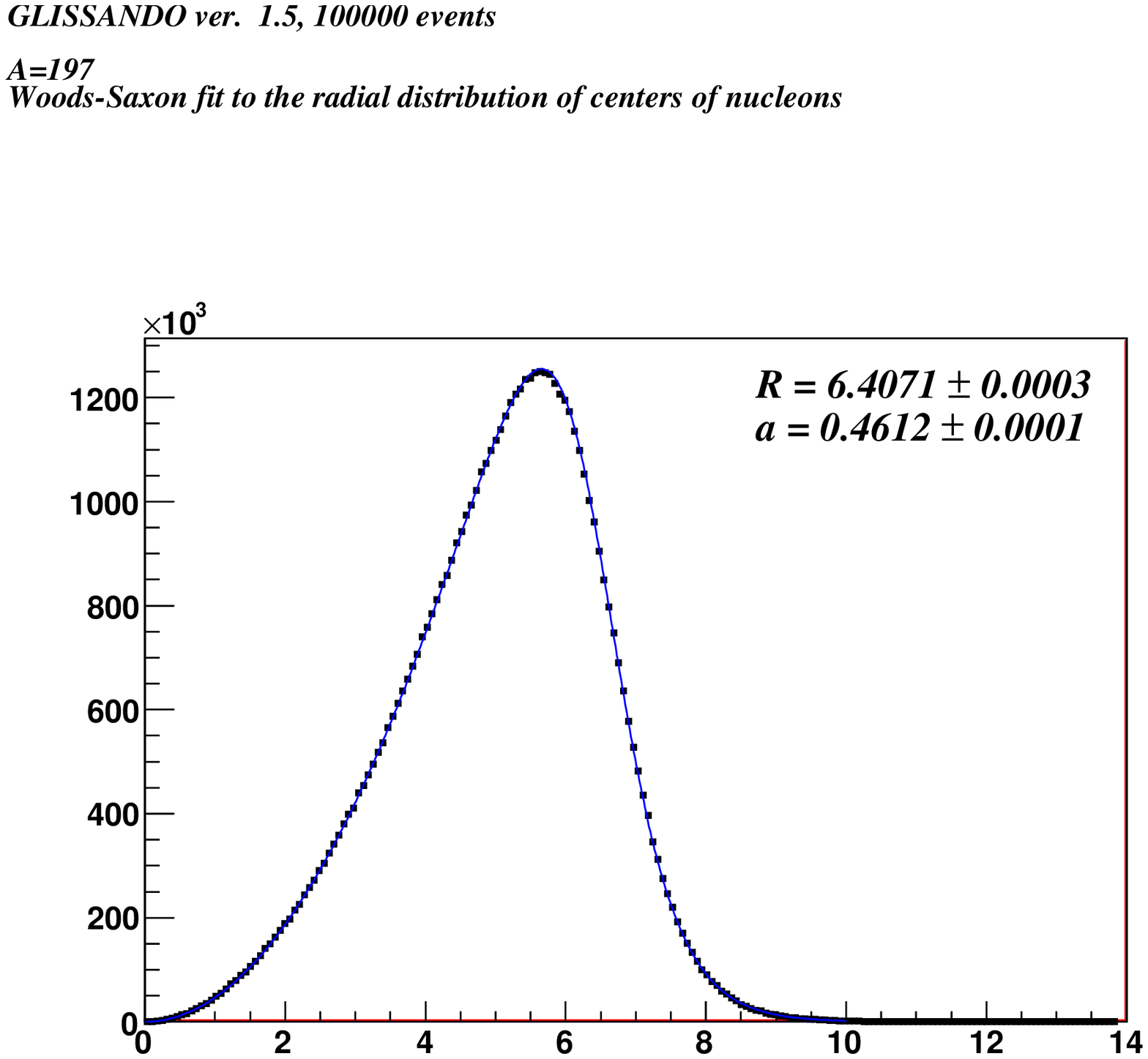}}\\
\subfigure{\includegraphics[width=.7\textwidth]{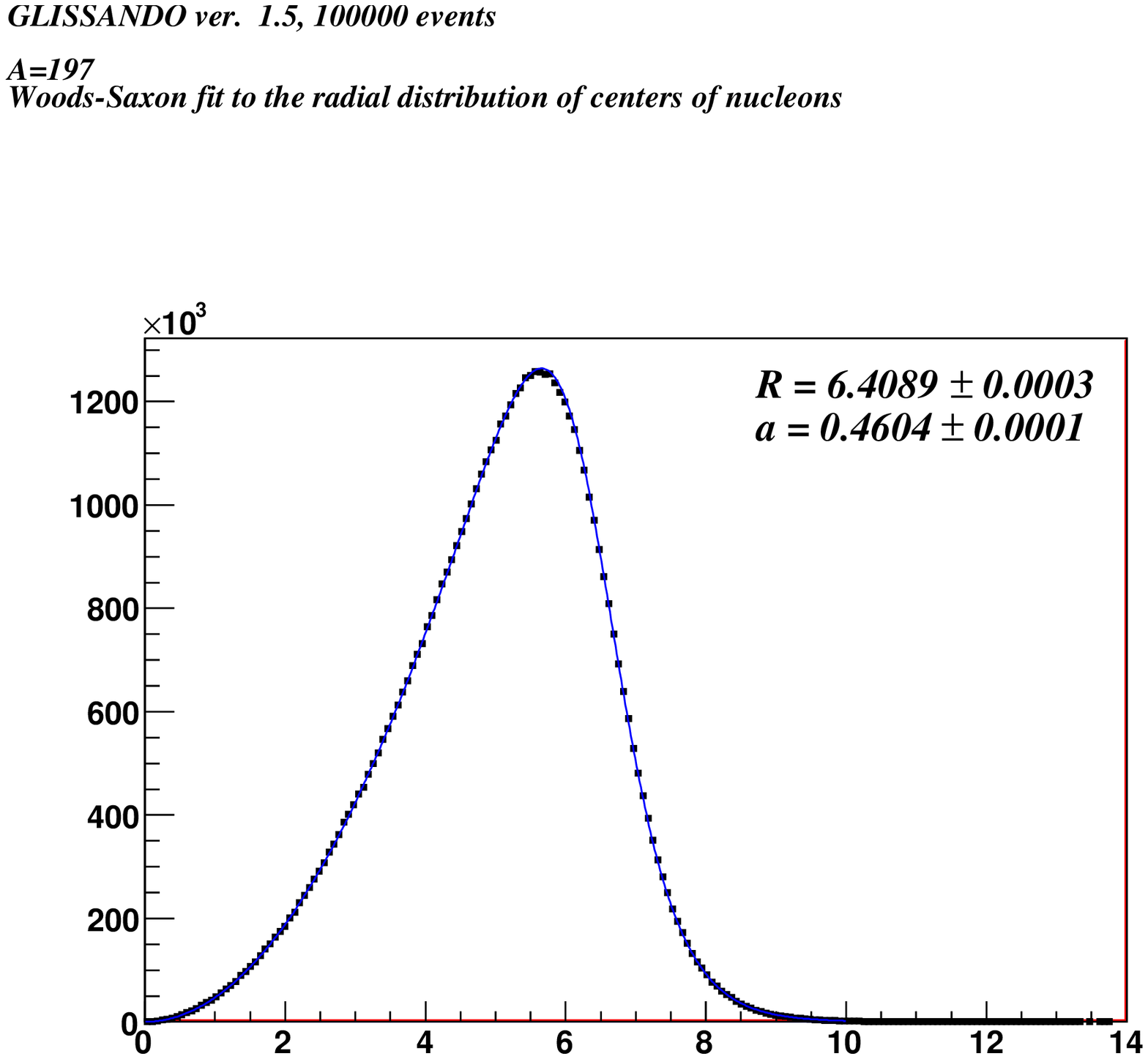}}
\end{center}
\caption{The histogram of the radial density for the distribution of centers of nucleons 
in the gold nucleus for the case of no expulsion distance (top), and 
with $d=0.4$~fm (bottom). Fixed-last algorithm, the bare parameters given in the text. \label{fig:fitr}}
\end{figure}

The user may first wish to verify the resulting nuclear density profile. For the case with no expulsion distance ($d=0$~fm) 
simply run

\begin{verbatim}
./glissando_profile.exe input_profile_0.dat
\end{verbatim}

\noindent
The file {\tt input\_profile\_0.dat} sets the bare (as explained in Sect.~\ref{sec:pos}) 
Woods-Saxon parameters for the distribution of the 
nucleon centers to $R=6.44$~fm, $a=0.45$~fm (gold) and uses no 
short-range expulsion ($d=0$~fm). The code generates the {\tt ROOT} output file {\tt glissando.root}. The user should then
enter the {\tt ROOT} interpreter

\begin{figure}[tb]
\begin{center}
\includegraphics[width=1\textwidth]{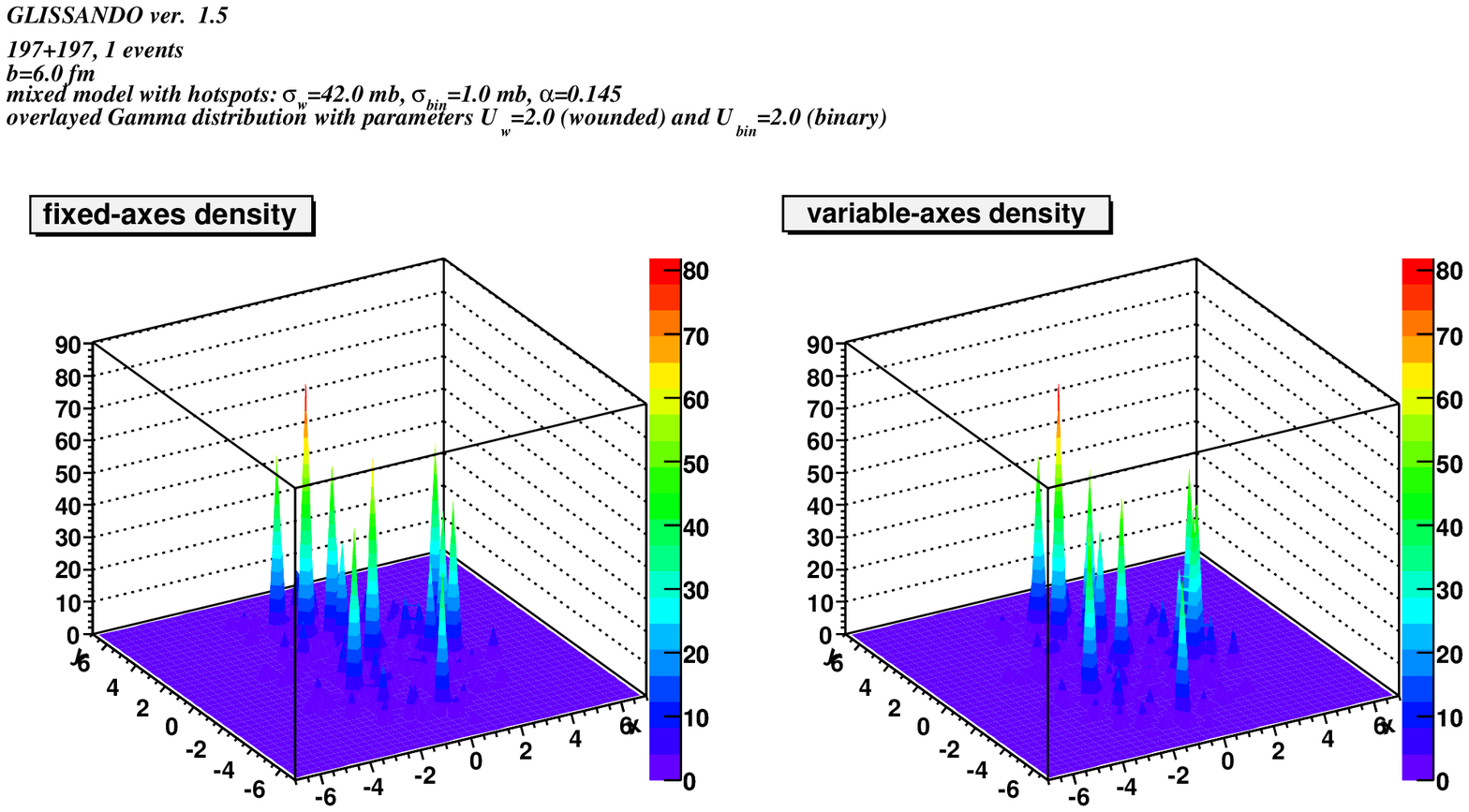}
\end{center}
\caption{The spatial distribution of RDS generated in the single event in the hot-spot model. \label{fig:hs}}
\end{figure}

\begin{verbatim}
root
\end{verbatim}

\noindent
and execute the script 

\begin{verbatim} 
.x fitr.C("")
\end{verbatim}

\noindent
The histogram of the nuclear density of nucleus $A$ stored in {\tt glissando.root} is fitted to the 
Woods-Saxon shape. The script generates the top panel of Fig.~\ref{fig:fitr} and displays the parameters $R$ and $a$, which 
are very close to the values of parameterization (\ref{rad0}).

For the case with expulsion distance of $d=0.4$~fm the ``bare'' Woods-Saxon parameters are $R=6.43$~fm, $a=0.45$~fm. 
The running sequence is as follows: 

\begin{verbatim}
./glissando_profile.exe input_profile_04.dat
root 
.x fitr.C("")
\end{verbatim}

\noindent 
The resulting plot and the fit parameters are shown in the bottom panel of Fig.~\ref{fig:fitr}.

In studies of reactions involving new nuclei
the user may tune the values of $R$ and $a$ in such a way, that with a chosen value of $d$ 
a proper density profile of nuclear centers is obtained. 

Now we are ready to run the simulations. First, let us look at a {\em single} event and make a snap-shot of the 
transverse-energy distribution in the $x-y$ plane. We use here for illustration the hot-spot model and fix the seed 
for the Monte Carlo generator, such the result is repeatable.

\begin{verbatim} 
./glissando.exe input_snap.dat
root 
.x density.C("")
\end{verbatim}

\noindent 
The produced plots are shown in Fig~\ref{fig:hs} (of course, for a single event the fixed- and variable-axes distributions are the 
same up to a rotation in the $x-y$ plane).

\begin{table}[tb]
\caption{Centrality classes in the impact parameter $b$, number of 
wounded nucleons $N_w$, and RDS for the mixed model 
with $\alpha=0.145$, gold-gold collisions. \label{tab:cent}} 
\begin{center}
{ 
\begin{verbatim}

b min: 0.0199137, b max: 19.3733
N_w min: 2, N_w max: 391
RDS min: 1, RDS max: 377.349

   c  --> b        N_w      RDS
0-5%  --> 3.10707, 326.834, 282.757
0-10% --> 4.59742, 277.708, 233.45
0-15% --> 5.44904, 236.603, 192.193
0-20% --> 6.30067, 200.51,  158.986
0-25% --> 7.1523,  170.433, 130.811
0-30% --> 7.79102, 142.361, 106.661
0-35% --> 8.42974, 118.299, 85.5293
0-40% --> 9.06847, 98.2481, 68.4228
0-45% --> 9.70719, 80.2018, 54.3351
0-50% --> 10.133,  64.1607, 42.2599
0-55% --> 10.5588, 51.1273, 32.1973
0-60% --> 10.9846, 39.0964, 24.1472
0-65% --> 11.6234, 30.0733, 18.1096
0-70% --> 12.0492, 22.0527, 13.0783
0-75% --> 12.475,  16.0373, 9.05324
0-80% --> 12.9008, 11.0244, 6.03445
0-85% --> 13.1137, 8.01671, 5.02819
0-90% --> 13.7524, 5.009,   3.01566
0-95% --> 14.3911, 3.00386, 2.0094

\end{verbatim}
}
\end{center}
\end{table}

\begin{figure}[tb]
\begin{center}
\subfigure{\includegraphics[width=.55\textwidth]{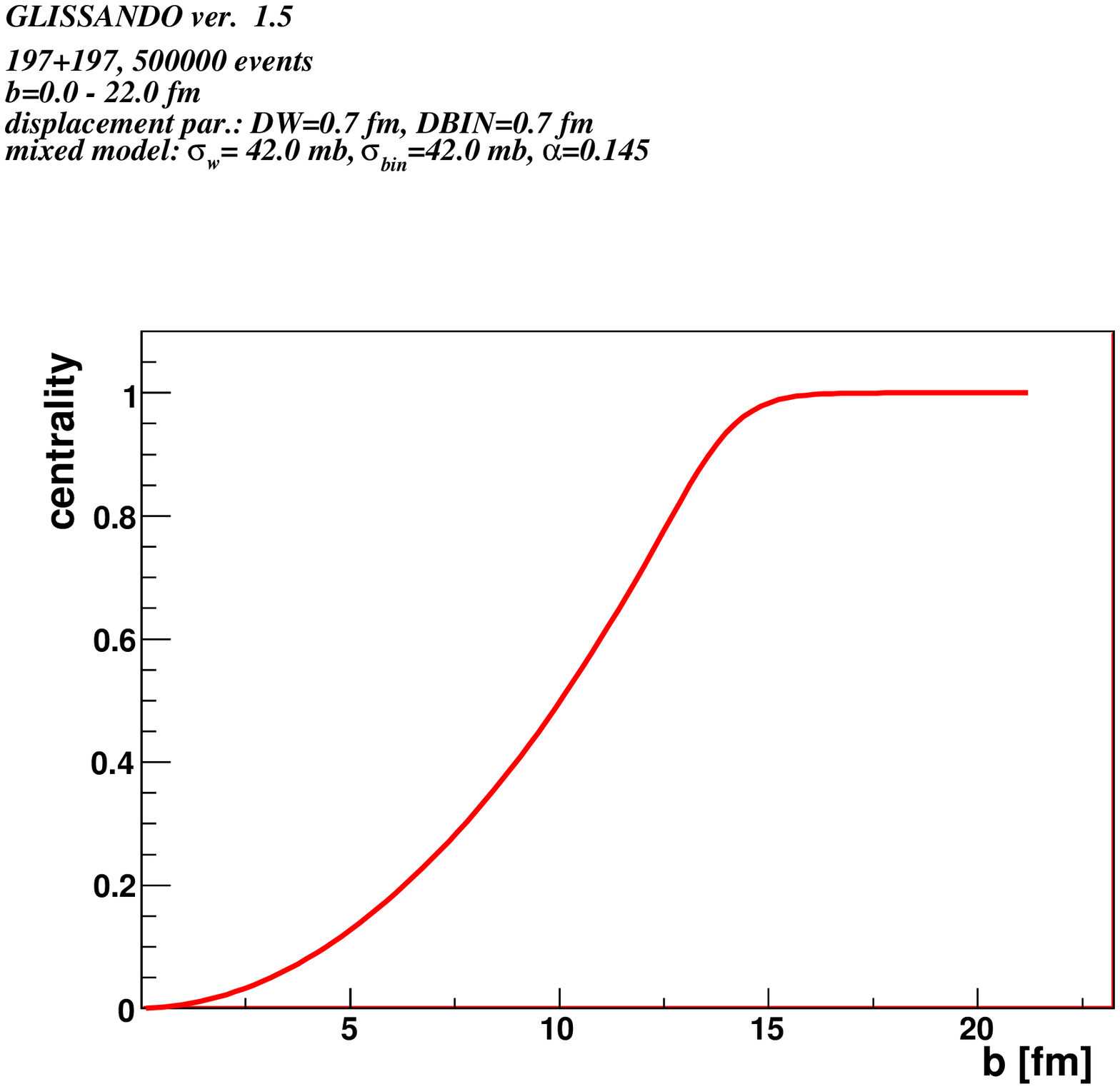}}\\
\subfigure{\includegraphics[width=.55\textwidth]{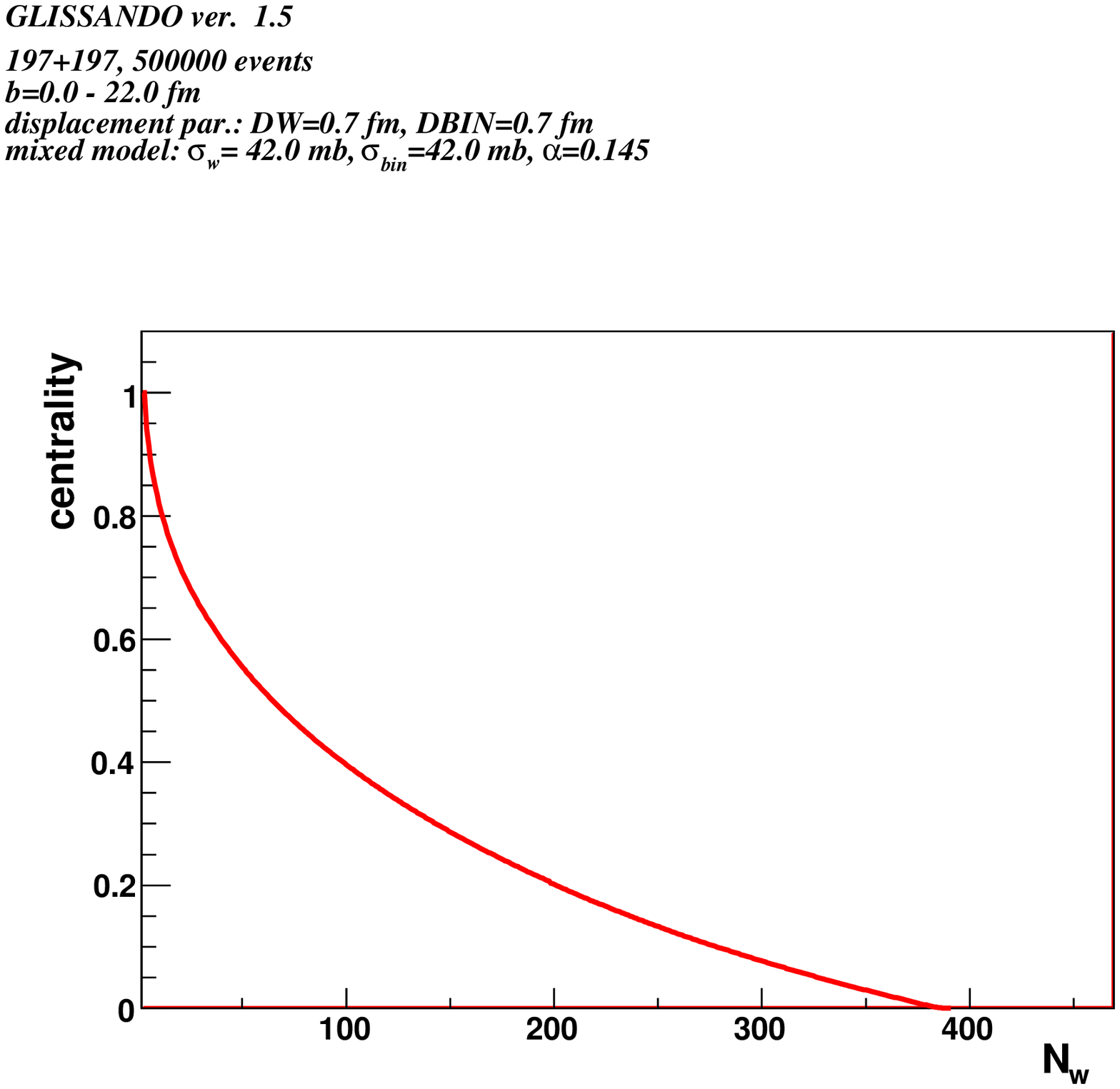}}\\
\subfigure{\includegraphics[width=.55\textwidth]{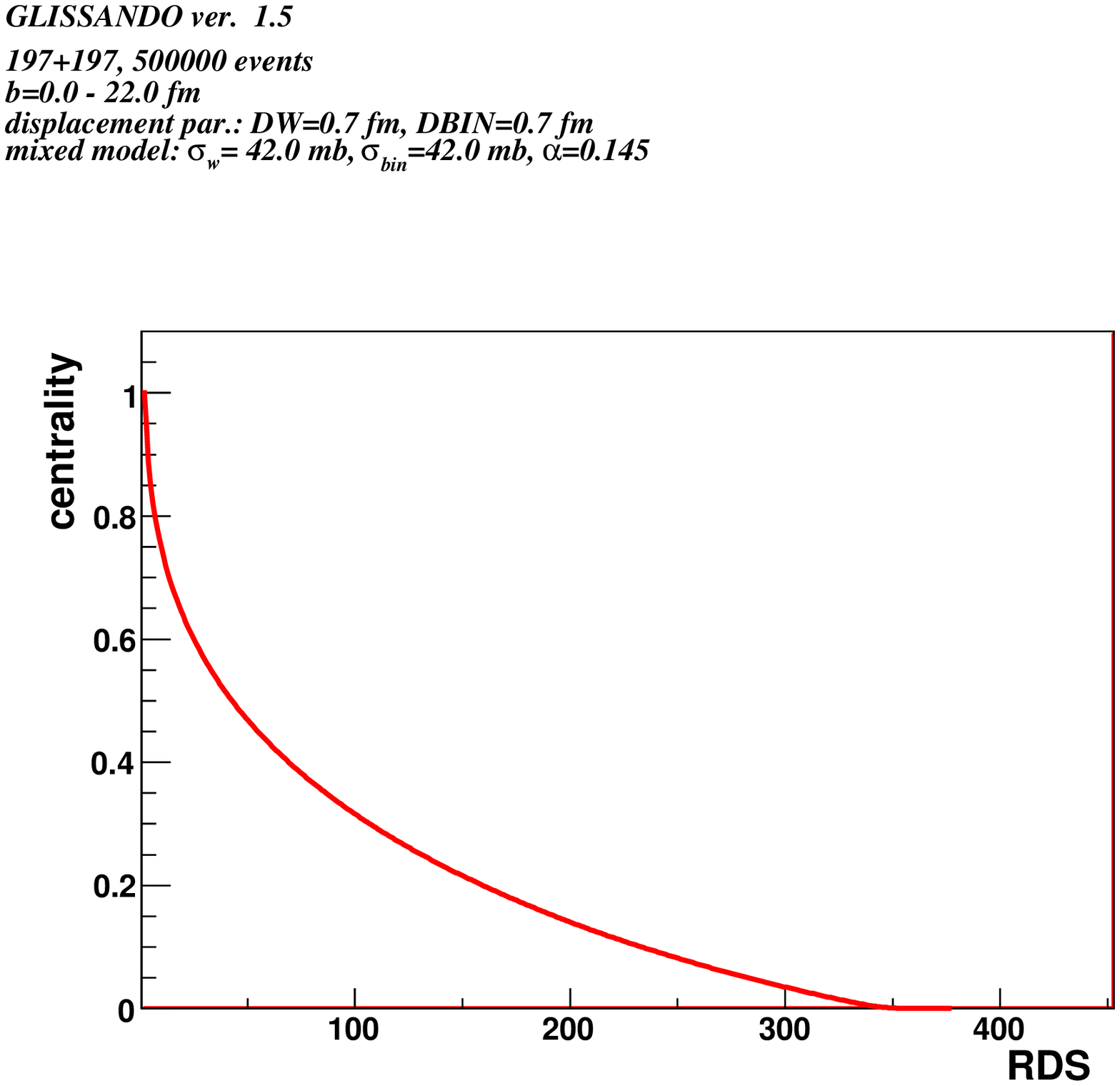}}
\end{center}
\caption{Centrality as the function of the impact parameter $b$, the number of wounded nucleons $N_w$, 
and the relative deposited strength (RDS). \label{fig:cent}}
\end{figure}

Next comes a full-fledged minimum-bias simulation for the {\em mixed} model for 
$Au$+$Au$ collisions with 500000 events: 

\begin{verbatim}
./glissando.exe input_minbias.dat
root 
.x centrality.C("")
\end{verbatim}

\noindent
The script {\tt centrality.C} determines the centrality classes in the $b$-parameter, in the number of 
wounded nucleons, $N_w$, and RDS. Since the number of produced particles is proportional to the RDS, 
the last case corresponds in essence to the determination of centrality via the multiplicity of produced particles.
The output of {\tt centrality.C} is generated to the console as well as to the 
file {\tt centrality.dat}. The file contains the information on the centrality classes in steps of 5\%,
as shown in Table~\ref{tab:cent}. Graphical representation of the results 
is shown in Fig.~\ref{fig:cent} generated by the script. 

A major interest has been recently attracted by the fixed- and variable-axes eccentricity, as its event-by-event 
fluctuations are linked to the fluctuations of the elliptic flow coefficient $v_2$. The following scripts show the dependence
of $\epsilon$, $\epsilon^\ast$, and their scaled standard deviations 
$\Delta \epsilon/\epsilon$ and $\Delta \epsilon^\ast/\epsilon^\ast$ as functions of $N_w$ and $b$: 

\begin{verbatim}
root
.x epsilon.C("")
.x epsilon_b.C("")
\end{verbatim}

\begin{figure}[tb]
\begin{center}
\subfigure{\includegraphics[width=.7\textwidth]{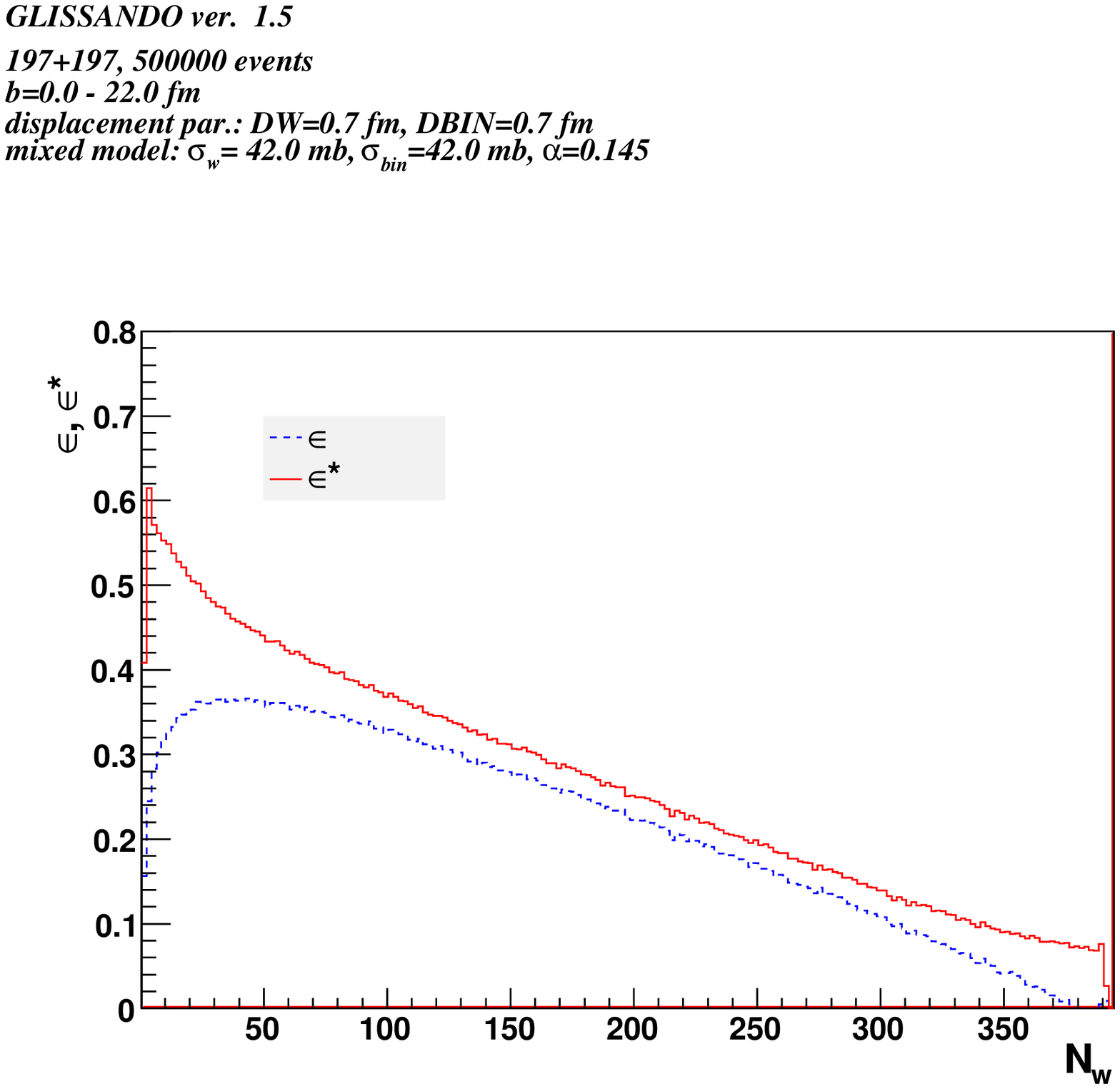}}\\
\subfigure{\includegraphics[width=.7\textwidth]{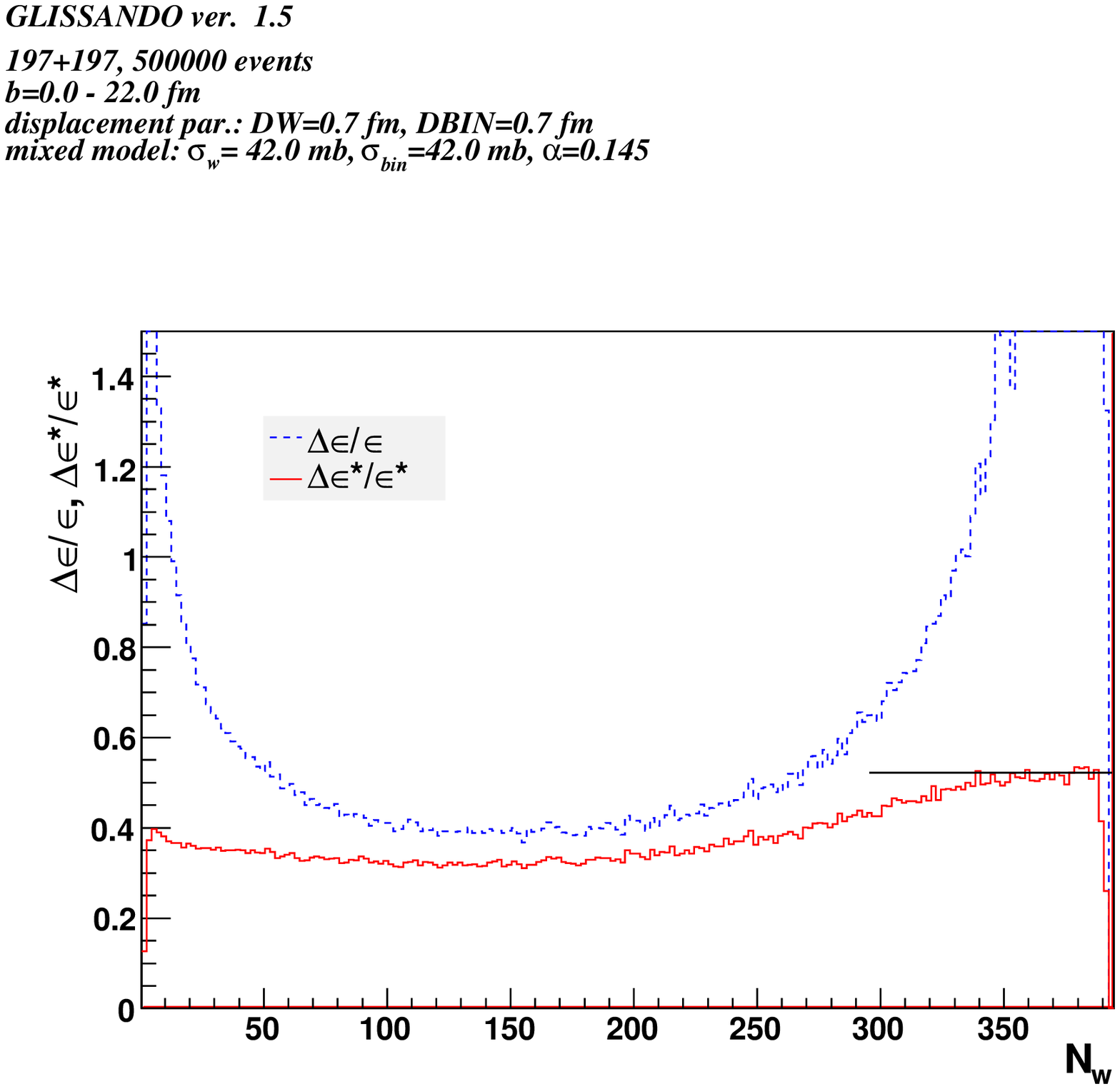}}
\end{center}
\caption{Fixed- and variable-axes eccentricities and their scaled standard deviations plotted as functions of $N_w$. 
The rebinning parameter was set to 1. \label{fig:eps}}
\end{figure}

\noindent The scripts ask for a rebinning parameter (a small natural number) which smooths out the plots. This is desired 
when the statistics is small. Originally, 200 bins for the impact parameter are used. If the rebinning parameter is
greater than 1, then this number of neighboring bins is grouped together. 
The result for the dependence on $N_w$ are shown in Fig.~\ref{fig:eps}. The horizontal solid line indicates the value
$\sqrt{4/\pi-1}\simeq 0.52$, which is the limit of $\Delta \epsilon^\ast/\epsilon^\ast$ for 
most central events\cite{Broniowski:2007ft}.

Similarly, the script {\tt dxdy.C} shows the dependence of the event-by-event standard deviation of the center-of-mass 
coordinates of the fireball on $N_w$:

\begin{verbatim}
root
.x dxdy.C("")
\end{verbatim}

We may now select the centrality window of interest and perform a simulation for that centrality class. 
For instance, for $c=0-5\%$ in $N_w$ we run

\begin{verbatim}
./glissando.exe input_0_5.dat glissando_0_5.root
\end{verbatim}

\noindent
(note that for better book keeping we now use a name for the {\tt ROOT} output file, which is stored for a potential later use).
The choice of the centrality in $N_w$ needs a careful selection of the window in $b$. Of course, one may use a
minimum-bias window such as $b=0-25$~fm, but this causes a waste of time, as many events are not recorded and the efficiency
is low. For the 
present case, we have found from Table~\ref{tab:cent} (or in {\tt centrality.dat}) 
that the selected centrality window corresponds to $N_w=327-394$. Thus, we 
have set {\tt W0=328} and {\tt W1=1000} (of course, any number $\ge 394$ is good) in the file {\tt input\_0\_5.dat}. 
It turns out that in this case it is enough to use $b=0-4.5$~fm. In order to verify if this selection is proper, 
we may perform a test by executing 

\begin{verbatim}
root
TBrowser a
\end{verbatim}

\noindent 
Selecting {\tt glissando\_0\_5.root} and clicking on the leaf $b$ generates a histogram of $b$. The 
supplied range in $b$ should be sufficiently wide, such that the histogram is not chopped off on the sides.

\begin{figure}[tb]
\begin{center}
\subfigure{\includegraphics[width=.72\textwidth]{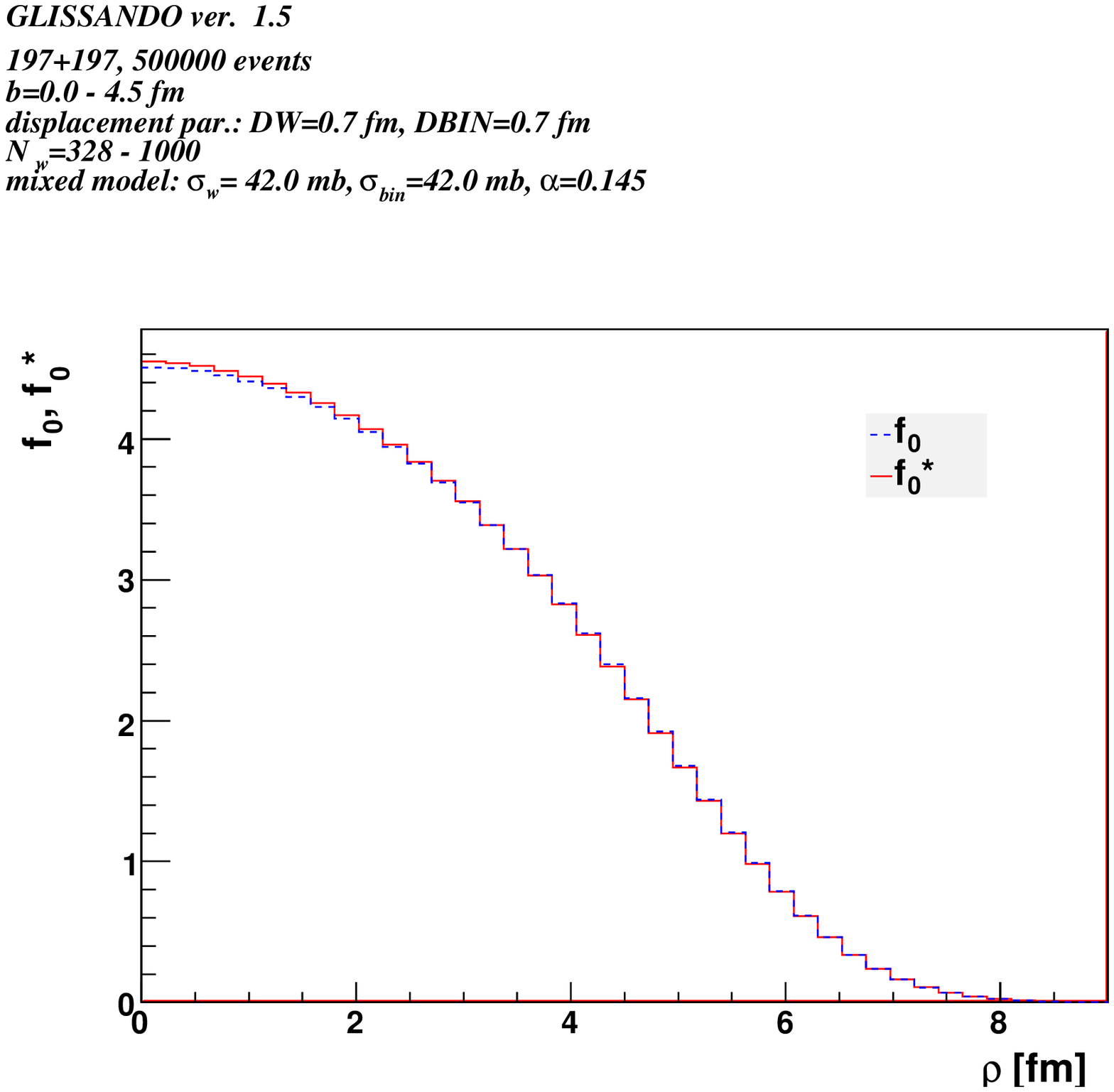}}\\
\subfigure{\includegraphics[width=.72\textwidth]{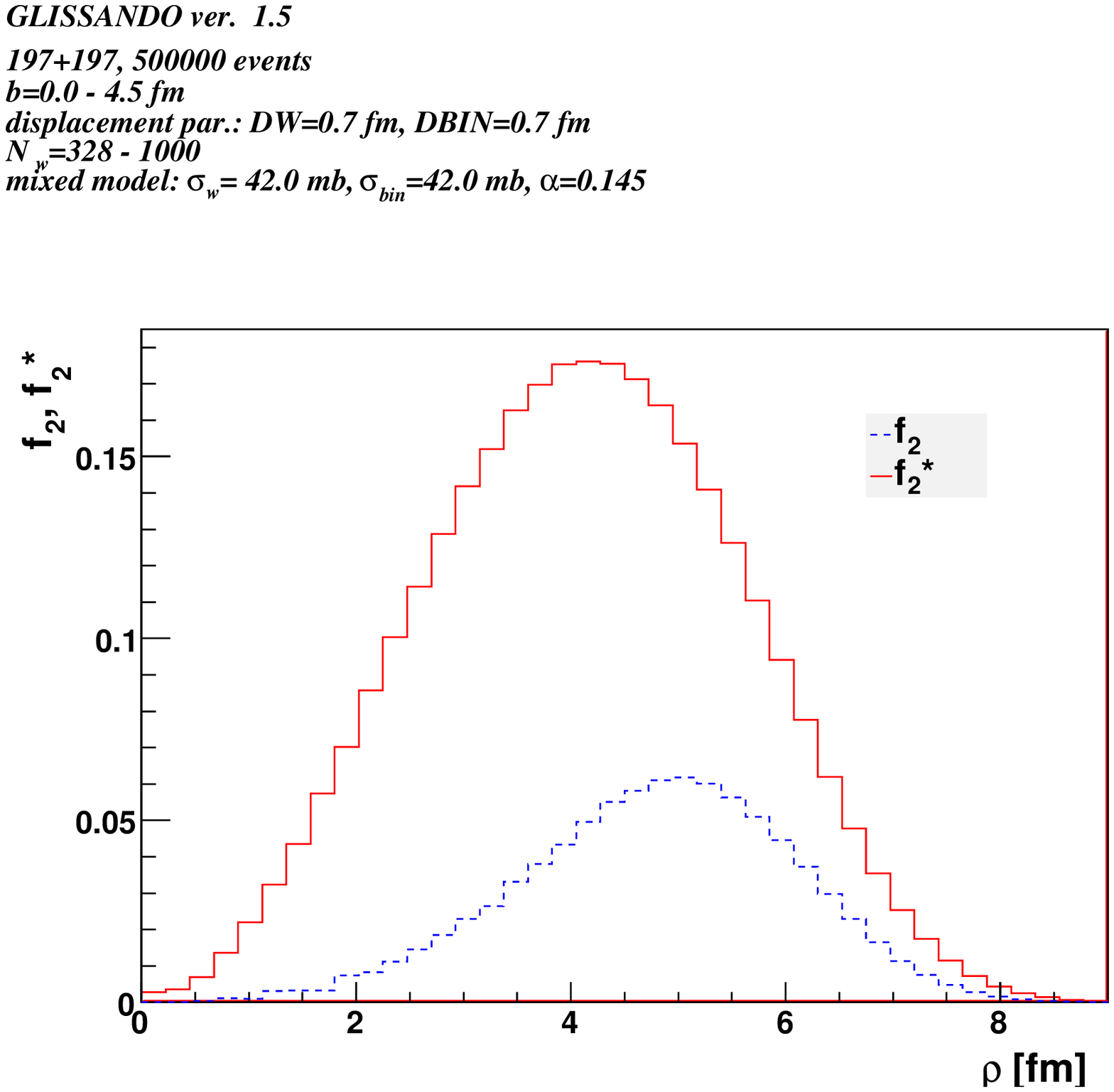}}
\end{center}
\caption{Fixed- and variable-axes Fourier profiles for $c=0-5\%$. \label{fig:prof}}
\end{figure}

Next, the scripts {\tt density.C} and {\tt profile2.C} generate plots of the two-dimensional fixed- 
and variable-axes profiles in the $x-y$ plane 
and their Fourier components. Note that large statistics is required for the higher Fourier profiles. 

\begin{verbatim}
root 
.x density.C("glissando_0_5.root")
.x profile2.C("glissando_0_5.root")
\end{verbatim}

\noindent 
An example is shown in Fig. \ref{fig:prof}, where we give the fixed- and variable-axes profiles for $l=0$ and $2$
harmonics.

The two-dimensional fireball 
profiles and their Fourier components may be used 
in further analyses, for instance in the calculation of jet
quenching, where we need the shape of the absorbing medium, in event-by-event hydrodynamics, where the initial conditions 
undergo fluctuation, in studies of the elliptic flow, {\em etc.} For that purpose we also provide the 
template {\tt C++} code {\tt interpolation.exe} which generates interpolated values of the functions from the stored two-dimensional 
or one-dimensional histograms. The typical running is 

\begin{verbatim}
./interpolation.exe glissando_0_5.root
Test of normalization
Mean RDS= 311.818
Integral over 2 Pi rho f0(rho) drho= 311.824
Integral over (xyhist) f(rho_x,rho_y) drho_x drho_y= 311.869
INTERPOLATION
ver 1.0

1 - One Dimensional Interpolation
2 - Two Dimensional Interpolation

Type 1 or 2 (any other - Exit)
1
1D Interpolation

Give name of 1D histogram
c0rhp
Give value of rho: 1.5
Interpolated value at 1.5 equals 4.31756
Once more?(y/n)
n
\end{verbatim}

\noindent The available names of one-dimensional histograms are
\begin{verbatim}
c0hp, c0rhp, c2hp, c2rhp, c4hp, c4rhp, c6hp, c6rhp, s3hp, s3rhp,
\end{verbatim}
where $c$ indicates the cosine, $s$ the sine moments (for unequal nuclei), 
the number 0, 2,\dots denotes the index of the Fourier harmonic, 
and $r$ labels the variable-axes profiles.
Similarly, the two-dimensional interpolation may be tested.
The names of histograms for the fixed- and variable-axes two-dimensional densities 
are: \begin{verbatim} xyhist, xyhistr. \end{verbatim}

\noindent The code serves as a template example for interfacing the {\tt GLISSANDO} results to other applications in {\tt C++}.

The analysis shown above may be repeated for other centralities, {\em e.g.}

\begin{verbatim}
./glissando.exe input_30_40.dat glissando_30_40.root
...
./glissando.exe input_80_95.dat glissando_80_95.root
\end{verbatim}

\begin{figure}[tb]
\begin{center}
{\includegraphics[width=.99\textwidth]{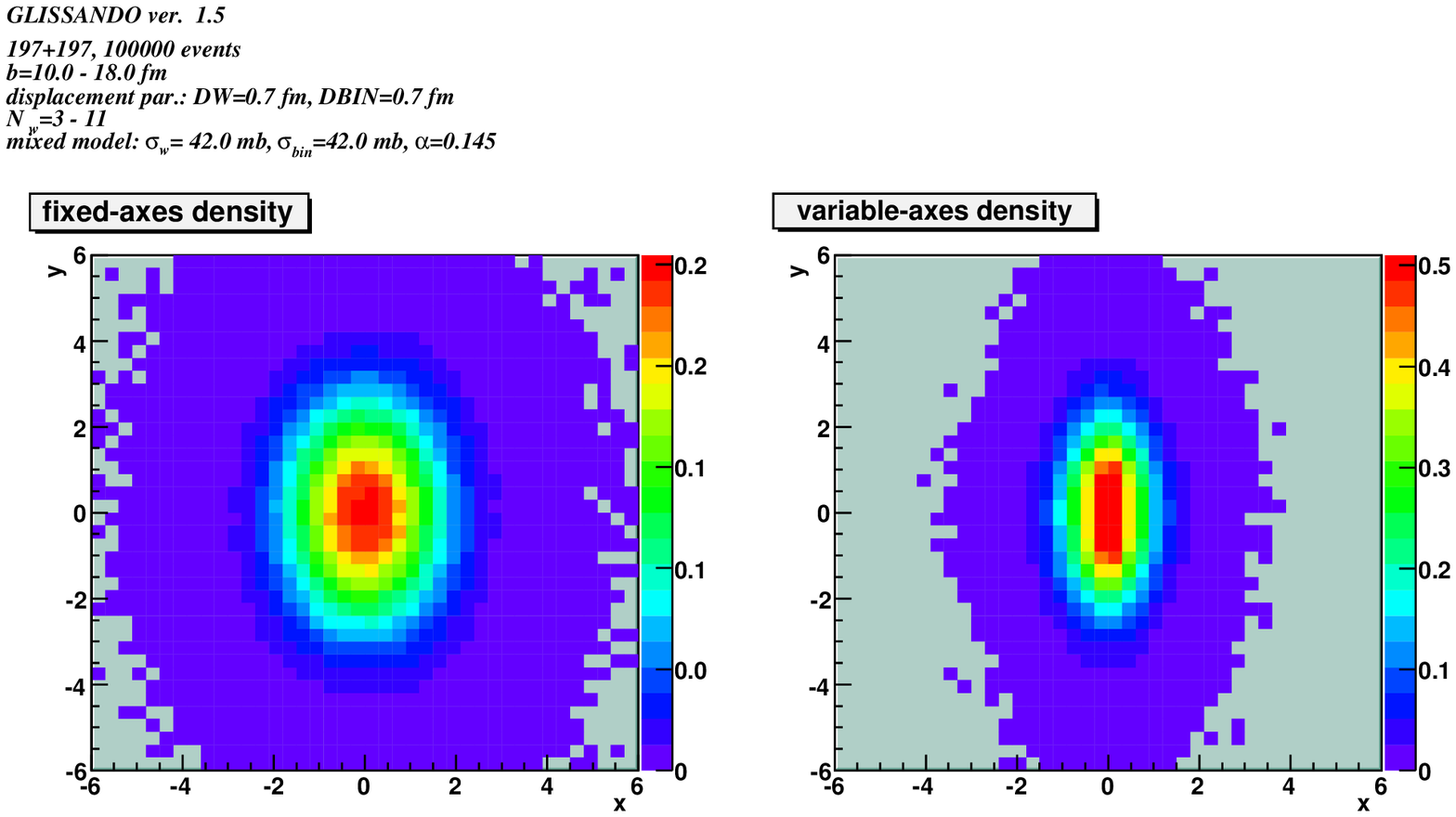}}\\
{\includegraphics[width=.99\textwidth]{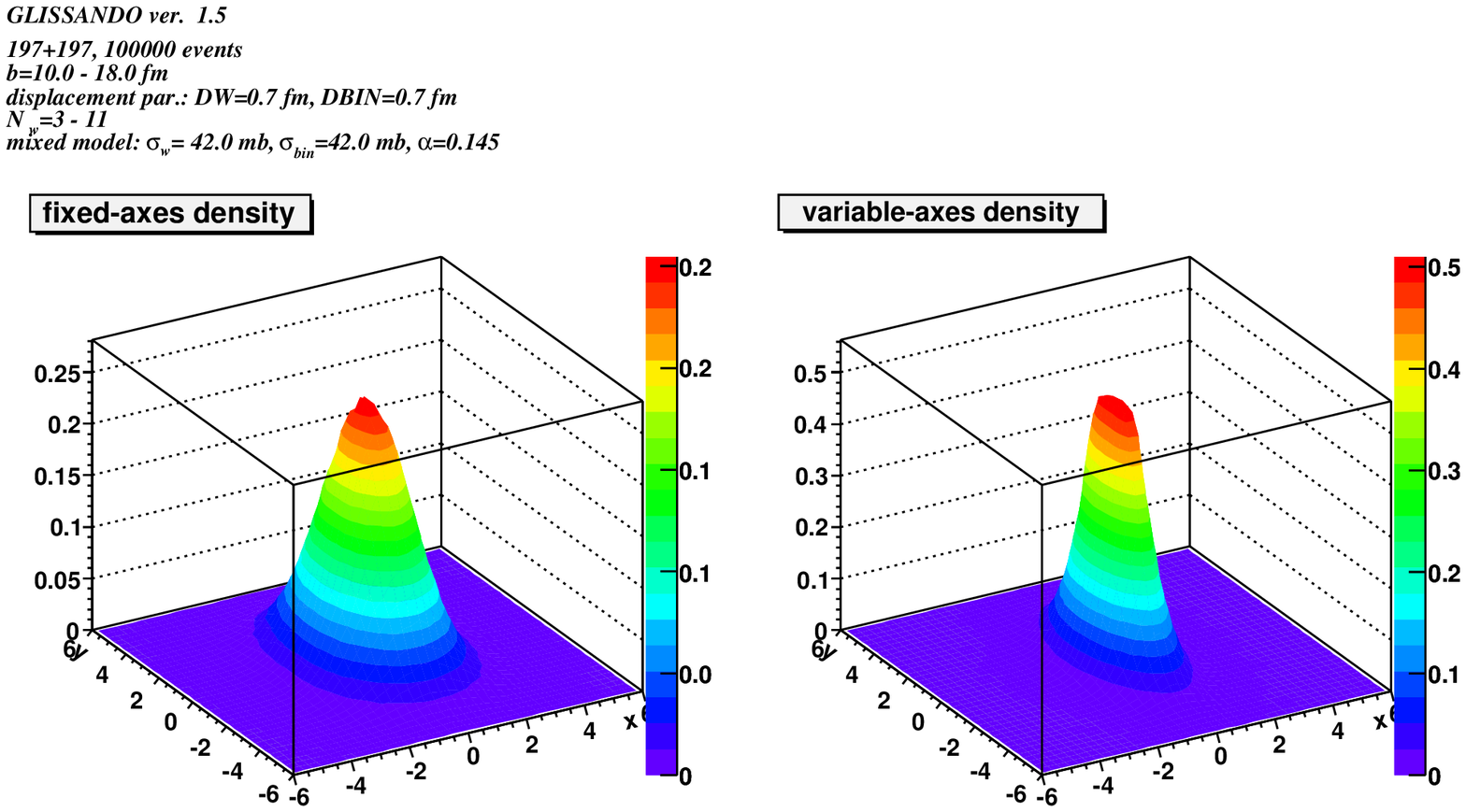}}\\
\end{center}
\caption{Three-dimensional density profiles for centrality 80-95\%.  \label{fig:den}}
\end{figure}

\noindent Figure~\ref{fig:den} shows the  two-dimensional profiles for $c=80-95\%$ obtained
with the command
\begin{verbatim}
root
.x density.C("glissando_80_95.root")
\end{verbatim}

We may carry out the calculation for other variants of the Glauber model, for instance for the hot-spot model
with superimposed $\Gamma$ distribution

\begin{verbatim}
./glissando.exe input_hotspot.dat hs.root
root 
.x centrality.C("hs.root")
.x epsilon.C("hs.root")
\end{verbatim}

\noindent
The code can be run for different nuclei (for the case of light nuclei the parameters of the 
nuclear distribution should be examined case by case, as the global parameterizations 
do not work accurately in this case). 

\begin{verbatim}
./glissando.exe input_S_Pb.dat
root 
.x centrality.C("")
.x epsilon.C("")
...
\end{verbatim}

\noindent
In the case of different nuclei it makes sense to shift the 
fireball to its center of mass also in the fixed-axes case, which is made by setting
{\tt SHIFT=1} in the input file.

For the proton-nucleus collisions we execute

\begin{verbatim}
./glissando_prot.exe input_prot.dat
root 
.x centrality.C("")
\end{verbatim}

\noindent
and for the deuteron-nucleus collisions

\begin{verbatim}
./glissando_deut.exe input_deut.dat
root 
.x centrality.C("")
\end{verbatim}

\noindent
We can obtain predictions for the upcoming LHC experiment by increasing the $\sigma_w$ 
cross section to $63$~mb (which is the 
appropriate value for the energy $\sqrt{s_{NN}}=5500$~GeV \cite{pdg})
and making an educated guess for the mixing parameter, $\alpha=0.2$:

\begin{verbatim}
./glissando.exe input_minbias_LHC.dat
root 
.x centrality.C("")
.x epsilon.C("")
\end{verbatim}

\section{Muliplicity fluctuations}

Histograms {\tt ntarg}, {\tt nbinar}, and {\tt nwei} in the {\tt GLISSANDO} output file contain, respectively, 
the average number of wounded nucleons in nucleus B, average number of binary collisions, and average RDS, obtained for a 
fixed number of 
wounded nucleons in nucleus A. Similarly, {\tt ntarg2}, {\tt nbinar2}, and {\tt nwei2} contain the scaled variances of these quantities. Such calculations are relevant for the multiplicity fluctuations at CERN SPS, where the 
number of wounded nucleons in nucleus $A$ (projectile) is measured with the help of the VETO calorimeter.

\section{Other results}

\begin{figure}[tb]
\begin{center}
\includegraphics[width=.8\textwidth]{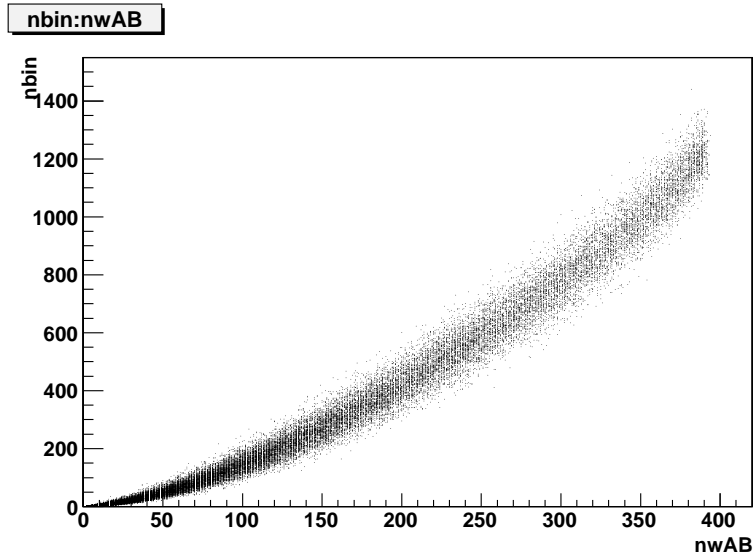}
\end{center}
\caption{Scattered plot for the correlation of numbers of wounded nucleons 
and binary collisions  for the minimum-bias gold-gold collisions at RHIC. \label{fig:wawb}}
\end{figure}

Numerous other results may be obtained interactively by 
entering the {\tt ROOT} interpreter and
accessing the {\tt GLISSANDO} output file.
As an example, we show in Fig.~\ref{fig:wawb} the 
event-by-event correlation plot between the number of wounded nucleons and the number of binary collisions
for the minimum-bias gold-gold collisions at RHIC (input file {\tt input\_minbias.dat}). The plot is generated by the following 
interactive sequence of instructions:

\begin{verbatim}
root
TBrowser a
[select the output file from GLISSANDO]
[select events tree]
[right-click on the events tree in the left window 
  and select Start Viewer]
[drop leaves nwAB and nbin on the x and y fields]
[hit draw from the bottom menu of the Viewer] 
\end{verbatim}

\section{Using the full event tree}

Setting the input parameter {\tt FULL} to 1 causes the generation of the full event tree, stored in the 
output {\tt GLISSANDO} file in the {\tt full\_event} tree. A sample run can be made as follows:
\begin{verbatim}
./glissando.exe input_full.dat
\end{verbatim}
The saved information contains the $x$ and $y$ coordinates 
({\tt X}, {\tt Y}) and the RDS ({\tt W}) of the source, as well as the event number ({\tt KK}). Technically, a structure is formed  
\begin{verbatim}
typedef struct {Float_t X,Y,W; UInt_t KK;} SOURCE;
static SOURCE tSource;
...
TTree *full_event; 
full_event = new TTree("full_event","full event");
full_event -> Branch("full_source",&tSource,"X:Y:W:KK");
\end{verbatim}
consecutively filled with information on each source, and written to the branch {\tt full\_source}. 
Other useful information, 
such as the impact parameter in the given event, is stored in the {\tt events} tree. A template use of the 
full event tree is shown in the program {\tt retrieve.exe}.
\begin{verbatim}
./retrieve.exe [glissando output file]
\end{verbatim}
The full event information may be processed ``off-line'' with other existing programs. 
The user should be warned that the FULL=1 option results in generating a lot of data, about 10MB/10000~events. 
Thus the option should only be used 
if the full results are to be used for analysis by another existing program.

\section{Summary}

We hope that with its flexibility and simplicity {\tt GLISSANDO} will become a useful tool for the heavy-ion 
community. The fact that the code is publicly available allows for check-ups, additions and improvements. We have provided 
examples of numerous applications: determination of the A+B cross-section and centrality classes, analysis
of eccentricities of the fireball
 both in the fixed- and variable axes frames, study of 
event-by-event fluctuations or correlation of various quantities.

\bigskip 
The authors are grateful to Adam Trzupek, Janusz Krywult, and Grzegorz Stefanek for useful comments, to Constantin Loizides
for helpful e-mail exchanges concerning the fluctuations of the center-of-mass of the fireball, and to Adam Bzdak for a discussion 
of the Gaussian wounding profile. We also thank Vittorio Soma for 
carrying out tests on the MacBook computer.

\newpage

\appendix

\section{Content of the package}

\begin{table*}[h]
\begin{tabular}{ll}
\hline
file name & description \\
\hline
\vspace{-2mm}
glissando.cpp & the {\tt GLISSANDO} source file \\\vspace{-2mm}
functions.cpp & the function source file\\\vspace{-2mm}
functions.h   & the functions header file \\\vspace{-2mm}
interpolation.cpp & template interpolation code source file \\\vspace{-2mm}
retrieve.cpp      & template code for retrieving info from the full event tree \\\vspace{-2mm} 
install & the installation shell script \\\vspace{-2mm}
Makefile & makefile for {\tt glissando.exe},  nucleus-nucleus collisions\\\vspace{-2mm}
Makefile.prot & makefile for {\tt glissando\_prot.exe}, proton-nucleus collisions\\\vspace{-2mm}
Makefile.deut & makefile for {\tt glissando\_deut.exe}, deuteron-nucleus collisions\\\vspace{-2mm}
Makefile.profile & makefile for {\tt glissando\_profile.exe}, profile for nucleus A\\\vspace{-2mm}
Makefile.inter & makefile for {\tt interpolation.exe} \\\vspace{-2mm}
Makefile.retr & makefile for {\tt retrieve.exe} \\\vspace{-2mm}
input.dat & the generic input \\\vspace{-2mm}
input\_profile\_0.dat & input for {\tt glissando\_profile.exe}, $d=0$~fm \\\vspace{-2mm}
input\_profile\_04.dat & input for {\tt glissando\_profile.exe}, $d=0.4$~fm \\\vspace{-2mm}
input\_snap.dat & input for generating a single event \\\vspace{-2mm}
input\_minbias.dat & input for minimum-bias gold-gold collisions at RHIC \\\vspace{-2mm}
input\_0\_5.dat     & input for centrality 0-5\%, gold-gold at RHIC \\\vspace{-2mm}
input\_30\_40.dat   & input for centrality 30-40\%, gold-gold at RHIC \\\vspace{-2mm}
input\_80\_95.dat   & input for centrality 80-95\%, gold-gold at RHIC \\\vspace{-2mm}
input\_S\_Pb.dat    & input for S-Pb collisions at SPS \\\vspace{-2mm}
input\_prot.dat    & input for proton-nucleus collisions \\\vspace{-2mm}
input\_deut.dat    & input for deuteron-nucleus collisions \\\vspace{-2mm}
input\_hotspot.dat   & input for the hot-spot model \\\vspace{-2mm}
input\_minbias\_LHC.dat   & input for LHC predictions for Pb-Pb collisions \\\vspace{-2mm}
input\_full.dat   & input generating the full event tree \\\vspace{-2mm}
info.C & script giving information on the stored output file\\\vspace{-2mm}
fitr.C & script for producing and fitting the nuclear profile\\\vspace{-2mm}
centrality.C & script for centrality classes\\\vspace{-2mm}
epsilon.C    & script for eccentricity vs. $N_w$, {\em etc.}\\\vspace{-2mm}
epsilon\_b.C  & script for eccentricity vs. $b$, {\em etc.}\\\vspace{-2mm}
dxdy.C       & script for center-of-mass coordinates vs. $N_w$\\\vspace{-2mm}
density.C    & script for two-dimensional profiles\\ \vspace{-2mm}
profile2.C   & script for Fourier profiles\\\vspace{-2mm}
label.C      & script generating the label used by other scripts\\\vspace{-2mm}
readme\_run   & short instructions for typical running\\\vspace{-2mm}
profile\_folding.nb & Mathematica file for folding the profile\vspace{-2mm}
\end{tabular}
\end{table*}

\section{Description of input and output \label{input}}

The basic model parameters, collected in Table~\ref{tab:input} can be supplied in the input file. The sign \# at the 
beginning of the line comments out the line and then the default value of the parameter is used.
See the sample file {\tt input.dat}. 

The variables stored in the output {\tt GLISSANDO} file are explained in Tables~\ref{tab:out1} and \ref{tab:out2}.

\begin{table*}[h]
\caption{Parameters of the input file. \label{tab:input}}
\begin{tabular}{lrl}
\hline
name & default & description \\
\hline \vspace{-3mm}
ISEED  & 0     &  seed for the random number generator, if 0 a random seed is generated \\\vspace{-3mm}
EVENTS & 50000 &  number of generated events \\\vspace{-3mm}
NBIN   & 40    &  number of bins for histograms in $\rho$, $x$, or $y$ \\\vspace{-3mm}
FBIN   & 72    &  number of bins for histograms in the azimuthal angle\\\vspace{-3mm}
NUMA   & 197   &  mass number of nucleus $A$ \\\vspace{-3mm}
NUMB   & 197   &  mass number of nucleus $B$ \\\vspace{-3mm}
RWSA   & 6.43  &  Woods-Saxon radius for the distribution of centers, nucleus $A$ [fm]\\\vspace{-3mm}
       &       &  gold with the fix-last method\\\vspace{-3mm}
AWSA   & 0.45  &  Woods-Saxon width, nucleus $A$ [fm]\\\vspace{-3mm}
RWSB   & 6.43  &  Woods-Saxon radius for the distribution of centers, nucleus $B$ [fm]\\\vspace{-3mm}
AWSB   & 0.45  &  Woods-Saxon width, nucleus $B$ [fm]\\\vspace{-3mm}
WFA    & 0     &  the w parameter of the Fermi distribution, nucleus $A$\\\vspace{-3mm}
WFB    & 0     &  the w parameter of the Fermi distribution, nucleus $B$\\\vspace{-3mm}
CD     & 0.4   &  closest allowed distance between centers of nucleons [fm]\\\vspace{-3mm}
SNN    & 42.   &  $NN$ ``wounding'' cross section [mb]\\\vspace{-3mm}
SBIN   & 42.   &  $NN$ binary cross section [mb]\\\vspace{-3mm}
ALPHA  & 0     &  0 - wounded, 1 - binary, 0.145 - PHOBOS for RHIC@200~GeV\\\vspace{-3mm}
MODEL  & 0     &  0 - constant superimposed weight=1, 1 - Poisson, 2 - Gamma\\\vspace{-3mm}
Uw     & 2     &  Poisson or Gamma parameter for wounded\\\vspace{-3mm}
Ubin   & 2     &  Poisson or Gamma parameter for binary\\\vspace{-3mm}
DW     & 0.7   &  dispersion of the location of the source for wounded nucleons [fm]\\\vspace{-3mm}
DBIN   & 0.7   &  dispersion of the location of the source for binary collisions [fm]\\\vspace{-3mm}
WMIN   & 2     &  minimum number of wounded nucleons to record event\\\vspace{-3mm}
W0     & 2     &  minimum allowed number of wounded nucleons\\\vspace{-3mm}
W1     & 1000  &  maximum allowed number of wounded nucleons\\\vspace{-3mm}
RDS0   & 0     &  minimum allowed RDS\\\vspace{-3mm}
RDS1   & 100000 &  maximum allowed RDS\\\vspace{-3mm}
GA     & 0.92  &  central value of the Gaussian wounding profile\\\vspace{-3mm}
SHIFT  & 0     &  1 - shift the coordinates of the fireball to 
                the c.m. in the fixed-axes case \\\vspace{-3mm}
       &       &  0 - do not shift\\\vspace{-3mm}
RET    & 0     &  0 - fix-last algorithm \\ \vspace{-3mm}
       &       &  1 - return-to-beginning algorithm for nuclear density\\\vspace{-3mm}
FULL   & 0     &  1 - provide the full information on events (coordinates and RDS of sources)\\\vspace{-3mm}
       &       &  0 - do not\\\vspace{-3mm}
DOBIN  & 0     &  1 - compute the binary collisions also for the case ALPHA=0 \\\vspace{-3mm}
       &       &  0 - do not  compute the binary collisions for the case ALPHA=0 \\\vspace{-3mm}
BMIN   & 0.    &  minimum impact parameter [fm]\\\vspace{-3mm}
BMAX   & 25.   &  maximum impact parameter [fm]\\
BTOT   &   &  range parameter for histograms [fm]\\
\hline
\end{tabular} 
\end{table*}

\begin{table*}[h]
\caption{Histograms stored in the output {\tt ROOT} file. $\langle . \rangle$ denotes the mean and var the variance 
of the specified quantity. \label{tab:out1}}
\begin{tabular}{ll}
\hline
\vspace{-3mm}
xyhist  & fixed-axes density in the $x-y$ variables\\ \vspace{-1mm}
xyhistr & variable-axes density in the $x-y$ variables\\ \vspace{-3mm}
c0hist  & fixed-axes density in the $\rho-\phi$ variables (not normalized)\\\vspace{-3mm}
c0rhist & variable-axes density in the $\rho-\phi$ variables (not normalized)\\\vspace{-3mm}
c0hp  & $f_0$\\ \vspace{-3mm}
c2hp  & $f_2$ - fixed-axes cosine harmonics \\ \vspace{-3mm}
c4hp  & $f_4$\\ \vspace{-3mm}
c6hp  & $f_6$\\ \vspace{-3mm}
c0rhp & $f^*_0$\\ \vspace{-3mm}
c2rhp & $f^*_2$ - variable-axes cosine harmonics\\ \vspace{-3mm}
c4rhp & $f^*_4$\\ \vspace{-3mm}
c6rhp & $f^*_6$\\ \vspace{-3mm}
s3hp  & $g_3$ - fixed-axes sine harmonic \\ \vspace{-3mm}
s3rhp & $g^*_3$ - variable-axes sine harmonic\\ \vspace{-3mm}
nx   &    $\langle x \rangle$ [fm] vs. $N_w$\\ \vspace{-3mm}
nx2  &    ${\rm var}(x)$ [fm]$^2$ vs. $N_w$\\ \vspace{-3mm}
ny   &    $\langle y \rangle$ [fm] vs. $N_w$\\ \vspace{-3mm}
ny2  &    ${\rm var}(y)$ [fm]$^2$ vs. $N_w$\\ \vspace{-3mm}
neps &    $\langle \epsilon \rangle$ vs. $N_w$\\ \vspace{-3mm}
neps2 &   ${\rm var}(\epsilon)/\langle \epsilon \rangle^2$ vs. $N_w$\\ \vspace{-3mm}
nepsp &   $\langle \epsilon^* \rangle$ vs. $N_w$\\ \vspace{-3mm}
nepsp2 &  ${\rm var}(\epsilon^*)/\langle \epsilon^* \rangle^2$ vs. $N_w$\\ \vspace{-3mm}
nuni   &  event multiplicity vs. $N_w$\\ \vspace{-3mm} 
nepsb &    $\langle \epsilon \rangle $ vs. $b$\\ \vspace{-3mm}
neps2b &   ${\rm var}(\epsilon)/\langle \epsilon \rangle^2$ vs. $b$\\ \vspace{-3mm}
nepspb &   $\langle \epsilon^* \rangle $ vs. $b$\\ \vspace{-3mm}
nepsp2b &  ${\rm var}(\epsilon^*)/\langle \epsilon^* \rangle^2$ vs. $b$\\ \vspace{-3mm}
nunib   &  event multiplicity vs. $b$\\ \vspace{-3mm}
ntarg  &   $\langle N_w^B \rangle $ vs. $N_w^A$\\ \vspace{-3mm}
ntarg2 &   ${\rm var}(N_w^B)/\langle N_w^B \rangle$ vs. $N_w^A$\\ \vspace{-3mm}
nbinar  &   $\langle N_{\rm bin} \rangle$ vs. $N_w^A$\\ \vspace{-3mm}
nbinar2 &   ${\rm var}(N_{\rm bin})/\langle N_{\rm bin} \rangle$ vs. $N_w^A$\\ \vspace{-3mm}
nwei  &   $\langle RDS \rangle$ vs. $N_w^A$\\ \vspace{-3mm}
nwei2 &   ${\rm var}(RDS)/\langle RDS \rangle$ vs. $N_w^A$\\ \\ \hline
\end{tabular}
\end{table*}

\begin{table*}[h]
\caption{Trees and their contents stored in the output {\tt ROOT} file.\label{tab:out2}}
\begin{tabular}{ll}
\hline
TTree param &  all parameters of the calculation \\
\hline
TTree density &  (generated only by {\tt glissando\_profile.exe})\\ \vspace{-3mm} 
r  &   radius of the nucleon in nucleus $A$ [fm]\\
wd &   weight generated by the superposition distribution\\\hline
TTree phys & \\\vspace{-3mm}
sitot & the nucleus-nucleus cross section [mb]\\\vspace{-3mm}
eps\_fixed & event-by-event average $\epsilon$  \\\vspace{-3mm}
eps\_variable & event-by-event average $\epsilon^\ast$ \\\vspace{-3mm}
sigma\_eps\_fixed & event-by-event standard deviation of $\epsilon$ \\
sigma\_eps\_variable & event-by-event standard deviation of $\epsilon^\ast$ \\
\hline
\end{tabular}
%
\begin{tabular}{ll}  
TTree events & \\ \vspace{-3mm}
nwA  & number of wounded nucleons in A\\ \vspace{-3mm}
nwB  & number of wounded nucleons in B\\ \vspace{-3mm}
nwAB & total number of wounded nucleons\\ \vspace{-3mm}
nbin & number of binary collisions\\ \vspace{-3mm}
npa  & RDS\\  \vspace{-1mm}
b    & impact parameter\\  \vspace{-3mm}
es   & fixed-axes epsilon (standard), $< r^2 \cos(2 \phi) >$ \\  \vspace{-3mm}
ep   & variable-axes epsilon (participant), $< r^2 \cos(2 (\phi-\phi^*)) >$ \\  \vspace{-3mm}
es3   & $< r^2 \cos(3 \phi) >$\\ \vspace{-3mm}
es4  & $< r^2 \cos(4 \phi) >$\\ \vspace{-3mm}
ep4  & $< r^2 \cos(4 (\phi-\phi^*)) >$\\ \vspace{-3mm}
es6  & $< r^2 \cos(6 \phi) >$\\ \vspace{-3mm}
ep6  & $< r^2 \cos(6 (\phi-\phi^*)) >$\\ \vspace{-3mm}
es3s & $< r^2 \sin(3 \phi) >$\\ \vspace{-1mm}
phir & the rotation angle $\phi^*$\\ \vspace{-3mm}
xx   & x c.m. coordinate [fm]\\ \vspace{-3mm}
yy   & y c.m. coordinate [fm]\\ \vspace{-3mm}\\ \hline
\end{tabular}
\end{table*}

\section{Description of the {\tt ROOT} scripts \label{scripts}}

Each script is called with the name of the {\tt ROOT} file generated by {\tt GLISSANDO}, 

\begin{verbatim}
root
.x  <script>.C("[{\tt GLISSANDO} file]")}
\end{verbatim}
\noindent The empty string in the argument of {\tt <script>.C("")} is equivalent to the default {\tt <script>.C("glissando.root")}.
\begin{verbatim}
info.C("[output ROOT file]")
\end{verbatim}
\noindent produces info on the stored {\tt ROOT} file.

\begin{verbatim}
fitr.C("[output ROOT file]")
\end{verbatim}
\noindent produces the density profile of centers of nucleons and its Woods-Saxon fit. The script works on {\tt ROOT} files 
generated with {\tt glissando\_profile.exe}, which contain the {\tt tree1} tree an the leaf {\tt r}.

\begin{verbatim}
centrality.C("[output ROOT file]")
\end{verbatim}
\noindent produces centrality windows for the stored {\tt ROOT} file. This makes sense for minimum-bias calculations.

\begin{verbatim}
epsilon.C("[output ROOT file]")
\end{verbatim}
\noindent produces plots of the mean and scaled standard deviation of
the deformation parameters $\epsilon$ and $\epsilon^\ast$ as functions of $N_w$. The script 
asks for a rebinning parameter (a natural number).

\begin{verbatim}
epsilon_b.C("[output ROOT file]")
\end{verbatim}
\noindent same as {\tt epsilon.C} but with $N_w$ replaced by the impact parameter $b$.

\begin{verbatim}
dxdy.C("[output ROOT file]")
\end{verbatim}
\noindent produces the event-by-event 
standard deviation of the coordinates of the center-of-mass of the fireball as a function of $N_w$.

\begin{verbatim}
density.C("[output ROOT file]")
\end{verbatim}
\noindent produces plots of the two-dimensional fixed-axes and variable-axes spatial distribution of RDS.

\begin{verbatim}
profile2.C("[output ROOT file]")
\end{verbatim}
\noindent produces plots of the fixed-axes and variable axes Fourier profiles of spatial distribution of RDS.


\end{document}